\documentclass[floatfix,twocolumn,showpacs,prd,aps,tightenlines,superscriptaddress]{revtex4-1}
\usepackage{graphicx}
\usepackage{enumerate}
\usepackage{color}
\usepackage{psfrag}
\usepackage{dcolumn}
\usepackage{bm}

\usepackage{amssymb}
\usepackage{amsmath}
\usepackage{amsfonts}
\usepackage{longtable}
\usepackage{xspace}

\usepackage{array}
\newcolumntype{H}{>{\lrbox0}c<{\endlrbox}@{}}

\def\eatcell#1\unskip{}
\newcolumntype{E}{>{\eatcell}c@{}}

\usepackage{collcell}
\makeatletter
\newcolumntype{G}{>{\collectcell\@gobble}c<{\endcollectcell}@{}}
\makeatother

\newcommand{\xsede}{{\sc Xsede}\xspace}

\newcommand{\beq}{\begin{equation}}
\newcommand{\eeq}{\end{equation}}

\newcommand{\be}{\begin{equation}}
\newcommand{\ee}{\end{equation}}
\newcommand{\bea}{\begin{eqnarray}}
\newcommand{\eea}{\end{eqnarray}}
\newcommand{\bes}{\begin{subequations}}
\newcommand{\ees}{\end{subequations}}

\usepackage{bm}

\begin{document}

\title{Post-Newtonian Quasicircular Initial Orbits for Numerical Relativity}

\author{James Healy}
\affiliation{Center for Computational Relativity and Gravitation,
School of Mathematical Sciences,
Rochester Institute of Technology, 85 Lomb Memorial Drive, Rochester,
New York 14623}

\author{Carlos O. Lousto}
\affiliation{Center for Computational Relativity and Gravitation,
School of Mathematical Sciences,
Rochester Institute of Technology, 85 Lomb Memorial Drive, Rochester,
New York 14623}

\author{Hiroyuki Nakano}
\affiliation{Faculty of Law, Ryukoku University, Kyoto 612-8577, Japan.}
\affiliation{Center for Computational Relativity and Gravitation,
School of Mathematical Sciences,
Rochester Institute of Technology, 85 Lomb Memorial Drive, Rochester,
New York 14623}
\affiliation{Department of Physics, Kyoto University, Kyoto 606-8502, Japan.}

\author{Yosef Zlochower}
\affiliation{Center for Computational Relativity and Gravitation,
School of Mathematical Sciences,
Rochester Institute of Technology, 85 Lomb Memorial Drive, Rochester,
New York 14623}

\date{\today}

\begin{abstract} 
  We use post-Newtonian (PN) approximations to determine the initial
  orbital and spin parameters of black hole binaries that lead to
  low-eccentricity inspirals when evolved with numerical relativity
  techniques. In particular, we seek initial configurations that lead
  to very small eccentricities at small separations, as is expected
  for astrophysical systems. We consider three cases: (i)
  quasicircular orbits with no radial velocity, (ii) quasicircular
  orbits with an initial radial velocity determined by radiation
  reaction, and (iii) parameters obtained form  evolution of the PN equations of motion from much
  larger separations.  We study eight cases of spinning,
  nonprecessing, unequal mass binaries. We then use several
  definitions of the eccentricity, based on orbital separations  and
  waveform phase and amplitude, and find that using the complete 3PN
  Hamiltonian for quasicircular orbits to obtain the tangential orbital
  momentum, and using the highest-known-order radiation reaction expressions to obtain the radial momentum,
  leads to the lowest eccentricity. The
  accuracy of this method even exceeds that of inspiral data based on
  3PN  and 4PN evolutions.
\end{abstract}

\pacs{04.25.dg, 04.25.Nx, 04.30.Db, 04.70.Bw} \maketitle

\section{Introduction}\label{sec:Intro}


With the 2005 breakthroughs~\cite{Pretorius:2005gq, Campanelli:2005dd,
Baker:2005vv}, numerical relativity techniques are now routinely used
to simulate the late inspiral and merger of black-hole binary systems
(BHB), and its detailed predictions of the gravitational waves (GWs)
produced by those systems have been recently observed by
LIGO~\cite{Abbott:2016blz,Abbott:2016nmj,Abbott:2016apu}.
Direct numerical solutions of Einstein's equations were used in 
the validation and interpretation of these discoveries 
\cite{Abbott:2016blz,TheLIGOScientific:2016wfe,TheLIGOScientific:2016src,Abbott:2016izl,Abbott:2016apu,Abbott:2016nmj,TheLIGOScientific:2016pea,Lovelace:2016uwp}.

Because LIGO is only sensitive to the last few orbits of
stellar-mass BHB mergers, and because of the large computational
resources needed to
produce these simulations, BHB simulations start at relatively
small initial separations in astrophysical terms. On the other hand,
by the time a BHB enters the LIGO sensitivity band we expect that
the eccentricity will be very
small because any eccentricity of astrophysical origin would have been
most likely
radiated away well before the late-inspiral at a rate proportional to
$d^{19/12}$~\cite{Peters:1964zz}, with $d$, the separation of the
binary (see, for instance, Fig.~6 of Ref.~\cite{Mroue:2010re} or
Fig.~9 in Ref.~\cite{Lousto:2015uwa}).

Given the current limitations of numerical relativity to carry out BHB
simulations down to merger (see though Ref.~\cite{Lousto:2013oza})
from initial separations much larger than $\sim
25M$~\cite{Lousto:2014ida, Szilagyi:2015rwa}, it is crucial to give
initial orbital parameters corresponding to an eccentricity
sufficiently low that its effect on the waveforms is below the
relevant accuracy requirements.

The problem of finding low-eccentricity parameters was recognized
early on. One powerful technique to generate these parameters uses
post-Newtonian (PN) quasicircular approximations to generate orbital parameters. 
The first systematic study (up to third order, i.e., 3PN)
for subsequent use in numerical
evolutions was performed in Ref.~\cite{Baker:2002qf}. The technique
that was used there to provide initial data is for the first time described
in detail in this current paper, as well as its completion to 3PN and
extension to incorporate 4PN corrections, in Sec.~\ref{sec:QC}.

Husa et al.~\cite{Husa:2007rh} pioneered an extension to this method.
Rather than using quasicircular parameters, they used the PN equations
of motion to evolve non-spinning binaries from large separations
($\sim 100M$) down to the separation of the start of the subsequent
numerical evolution ($\sim 10 M$). They then used the evolved PN
orbital parameters to construct initial data.  With this technique, they
were able to generate reasonably low eccentricity values.

Our group used a similar technique 
to provide low eccentricity initial data for spinning and precessing
binaries in Ref.~\cite{Campanelli:2008nk}
and in subsequent papers.

The most accurate way to produce low-eccentricity data is to use an
iterative procedure to correct the initial orbital parameters until
sufficiently low eccentricities are
obtained~\cite{Pfeiffer:2007yz, Buonanno:2010yk, Purrer:2012wy,
Buchman:2012dw}. Each
step in this iterative procedure requires the evolution of a binary
for a few orbits. Thus this requires several full numerical evolutions
before starting the definitive one.

In this paper, we revisit the scenario of quasicircular orbits
as defined by the PN approximation. We update our
previous approach from incomplete 3PN to complete 3PN,
and include the possibility
of an initial radial velocity based on instantaneous radiation
reaction terms. Finally, we also consider the direct evolution
of the spinning 3PN and 4PN equations of motion from large
separations down to the desired starting separation to 
provide initial parameters for the numerical
evolutions.

The paper is organized as follows. In Sec.~\ref{sec:PN},
we describe how to compute quasicircular orbits up to 3PN
and how to calculate the radial inspiral velocity based in
the radiation reaction terms. We also describe the
4PN equations of motion and how a PN inspiral simulation can
be used to provide the alternative initial
parameters. In Sec.~\ref{sec:FN}, we explore eight different
spinning (but non-precessing) BHB configurations with initial orbital
parameters given by incomplete 3PN and complete 3PN quasicircular parameters
(with and without an initial radial velocity), and
3PN and 4PN inspiral parameters (a total of 40 individual simulations).
We conclude with a discussion in 
Sec.~\ref{sec:Discussion} of the benefits of each method
to obtain a first and computationally simple set of initial
configurations leading to small eccentricity for applications
not requiring initial eccentricities below $\sim10^{-3}$.

Throughout this paper we use geometric units where $G=c=1$. The
vacuum general relativity field equations are scale invariant. 
In the case of a black hole binary, rescaling
the total mass while keeping the mass ratio fixed,
keeping the dimensionless spins fixed,
and rescaling the momenta by the same factor as the masses
leads to an equivalent solution. When
reporting quantities with dimension, we rescale each by an
appropriate power of an arbitrary positive constant $M$ (which has
dimensions of mass).

\section{Post-Newtonian orbits}\label{sec:PN}

\subsection{Quasicircular Initial Parameters}\label{sec:QC}

Our construction of quasicircular orbital parameters is based on
PN dynamics in the
 Arnowitt-Deser-Misner transverse traceless
 (ADM-TT) gauge. This gauge is closely related (up to 2PN)
 to isotropic gauge used
in puncture initial data~\cite{Kelly:2007uc,Tichy:2002ec}.
From the Hamiltonian, i.e., the conservative part of the
PN dynamics, we get the ADM mass $M_{\rm ADM}$, orbital separation
$r$, 
and tangential linear momentum $P_t$ all in terms of the orbital
frequency $\Omega$, or, alternatively, 
 $M_{\rm ADM}$,  $P_t$, and  $\Omega$ can be obtained as functions
 of $r$. The mass, separation, and linear momenta are all
calculated in the center of mass.
To get the locations of the two BHs, we also need expressions for the
position of the center of mass.
Finally, the dissipative part, i.e., the radiation reaction gives
the radial momentum $P_r$.

In an actual numerical simulation, we use the ADM-TT positions of the two
BHs, their tangential and radial momenta, and their dimensionless
spins to construct the initial
data. That is, we take these ADM-TT position and momentum parameters,
as well as the parameters $\vec S_i = m_i^2 \vec \chi_i$, where 
$m_i$ is the PN mass parameter and $\chi_i$ is the PN dimensionless
spin of particle $i$, and plug them directly
into the Brandt-Br\"ugmann initial 
puncture formalism~\cite{Brandt97b}. 
The puncture masses of the two BHs is set by
demanding that the numerical horizon (Christodoulou) masses
match the ADM-TT mass parameters.

\subsubsection{From the Hamiltonian}

We start from the 3PN ADM-TT Hamiltonian,
\bea
H &=& H_{\rm O,Newt} + H_{\rm O,1PN} + H_{\rm O,2PN} + H_{\rm O,3PN}
\cr && + H_{\rm SO,1.5PN} + H_{\rm S_1S_2,2PN} + H_{\rm S^2,2PN} 
\cr && +H_{\rm SO,2.5PN} + H_{\rm S_1S_2,3PN} + H_{\rm S^2,3PN} \,.
\label{eq:3PNH}
\eea
The terms in Eq.~(\ref{eq:3PNH}) were taken from
Ref.~\cite{Buonanno:2005xu} augmented with 
the next-to-leading order (NLO) spin-orbit coupling, $H_{\rm SO,2.5PN}$,
derived by Ref.~\cite{Damour:2007nc} 
and the NLO $S_1$-$S_2$ coupling, $H_{\rm S_1S_2,3PN}$,
derived by Refs.~\cite{Steinhoff:2007mb, Steinhoff:2008zr}
that was used in Ref.~\cite{Campanelli:2008nk}.
The above Hamiltonian also has the NLO spin-squared ($S_1^2$ and $S_2^2$) terms,
$H_{\rm S^2,3PN}$,
presented in Ref.~\cite{Steinhoff:2008ji}.

In standard spherical coordinates, $\{r,\,\theta,\,\phi\}$,
the quasicircular conditions
in absence of radiation reaction are
\bea
P_r = 0 \,, \quad
\frac{\partial H}{\partial r} = 0
\,.
\eea
With this information, we can solve for $P_{\phi}$ as a function of
$r$ and can then obtain the
initial orbital frequency
\bea
\Omega = \left(\frac{\partial H}{\partial P_{\phi}}\right) \,,
\eea
as a function of $r$, as well.
Note that since it is not necessary to solve for $\dot P_{\phi}$,
we do not need to evaluate $\partial H/\partial \phi$.
The tangential linear momentum is given by
\bea
P_t = \frac{P_{\phi}}{r} \,,
\eea
where $P_t=P_{1y}=-P_{2y}$ for the individual BHs (the BHs are assumed
to lie on the $x$-axis initially).
The ADM mass is given by
\bea
M_{\rm ADM} = M + H \,,
\eea
where $M=m_1+m_2$ is the total mass,
and the Hamiltonian $H$ is calculated with $P_r=0$ and the solution of $P_{\phi}$.
Finally, the total ADM angular momentum of this system is
$\vec{J}=\{S_{1x}+S_{2x},\,S_{1y}+S_{2y},\,L_z+S_{1z}+S_{2z}\}$
with $L_z=P_{\phi}$, and the BH's spins, $\vec{S}_1$ and $\vec{S}_2$.

In the following, we summarize some quantities derived from the above
analysis.
Here, we use the mass ratio, $q=m_1/m_2$,
the nondimensional spin, 
$\chi_{1x}=S_{1x}/m_1^2$, $\chi_{1y}=S_{1y}/m_1^2$,
$\chi_{1z}=S_{1z}/m_1^2$, $\chi_{2x}=S_{2x}/m_2^2$, $\chi_{2y}=S_{2y}/m_2^2$,
and $\chi_{2z}=S_{2z}/m_2^2$.
The symmetric mass ratio is given by $\eta=q/(1+q)^2$.
First, all quantities are written in terms of the orbital separation $r$.
The orbital frequency, tangential linear momentum, and the ADM mass 
are given by
\begin{widetext}
\bea
M \Omega &=&
\left(\frac{{M}}{r}\right)^{3/2}
\Biggl[1
-\frac{1}{2}\,{\frac { \left( 3\,{q}^{2}+5\,q+3 \right) }
{\left( 1+q \right) ^{2} }}
\frac{{M}}{r}
+ \left( -\frac{1}{4}\,{\frac { \left( 3+4\,q \right) q{\chi_{1z}}}
{\left( 1+q \right) ^{2}}}
-\frac{1}{4}\,{\frac { \left( 3\,q+4 \right) {\chi_{2z}}}
{\left( 1+q \right) ^{2}}} \right)
\left(\frac{{M}}{r}\right)^{3/2}
\cr &&
+ \left( -\frac{3}{2}\,{\frac {{{\chi_{1x}}}^{2}{q}^{2}}
{ \left( 1+q \right) ^{2}}}
-3\,{\frac {{\chi_{1x}}\,{\chi_{2x}}\,q}{ \left( 1+q \right) ^{2}}}
+\frac{3}{4}\,{\frac {{{\chi_{1y}}}^{2}{q}^{2}}{ \left( 1+q \right) ^{2}}}
+\frac{3}{2}\,{\frac {{\chi_{1y}}\,{\chi_{2y}}\,q}{ \left( 1+q \right) ^{2}}}
+\frac{3}{4}\,{\frac {{{\chi_{1z}}}^{2}{q}^{2}}{ \left( 1+q \right) ^{2}}}
+\frac{3}{2}\,{\frac {{\chi_{1z}}\,{\chi_{2z}}\,q}{ \left( 1+q \right) ^{2}}}
-\frac{3}{2}\,{\frac {{{\chi_{2x}}}^{2}}{ \left( 1+q \right) ^{2}}}
\right. \cr && \left.
+\frac{3}{4}\,{\frac {{{\chi_{2y}}}^{2}}{ \left( 1+q \right) ^{2}}}
+\frac{3}{4}\,{\frac {{{\chi_{2z}}}^{2}}{ \left( 1+q \right) ^{2}}}
+\frac{1}{16}\,{\frac {24\,{q}^{4}+103\,{q}^{3}+164\,{q}^{2}+103\,q+24}
{ \left( 1+q \right) ^{4}}} \right)
\left(\frac{{M}}{r}\right)^{2}
\cr &&
+ \left( \frac{3}{16}\,{\frac {q \left( 16\,{q}^{3}+30\,{q}^{2}+34\,q+13 \right) 
{\chi_{1z}}}{ \left( 1+q \right) ^{4}}}
+\frac{3}{16}\,{\frac { \left( 13\,{q}^{3}+34\,{q}^{2}+30\,q+16 \right) {\chi_{2z}}}
{ \left( 1+q \right) ^{4}}} \right)
\left(\frac{{M}}{r}\right)^{5/2}
\cr &&
+ \left( \frac{1}{16}\,{\frac { \left( 76\,{q}^{2}+180\,q+155 \right) {q}^{2}
{{\chi_{1x}}}^{2}}{ \left( 1+q \right) ^{4}}}
+\frac{1}{8}\,{\frac { \left( 120\,{q}^{2}+187\,q+120 \right) q{\chi_{2x}}
\,{\chi_{1x}}}{ \left( 1+q \right) ^{4}}}
-\frac{1}{8}\,{\frac { \left( 43\,{q}^{2}+85\,q+55 \right) {q}^{2}{{\chi_{1y}}}^{2}}
{ \left( 1+q \right) ^{4}}}
\right. \cr && \left.
-\frac{1}{4}\,{\frac { \left( 54\,{q}^{2}+95\,q+54 \right) q{\chi_{2y}}\,{\chi_{1y}}}
{ \left( 1+q \right) ^{4}}}
-\frac{1}{32}\,{\frac { \left( 2\,q+5 \right)  \left( 14\,q+27 \right) {q}^{2}
{{\chi_{1z}}}^{2}}{ \left( 1+q \right) ^{4}}}
-\frac{1}{16}\,{\frac { \left( 96\,{q}^{2}+127\,q+96 \right) q{\chi_{2z}}
\,{\chi_{1z}}}{ \left( 1+q \right) ^{4}}}
\right. \cr && \left.
+\frac{1}{16}\,{\frac { \left( 155\,{q}^{2}+180\,q+76 \right) 
{{\chi_{2x}}}^{2}}{ \left( 1+q \right) ^{4}}}
-\frac{1}{8}\,{\frac { \left( 55\,{q}^{2}+85\,q+43 \right) {{\chi_{2y}}}^{2}}
{ \left( 1+q \right) ^{4}}}
-\frac{1}{32}\,{\frac { \left( 27\,q+14 \right) \left( 5\,q+2 \right) 
{{\chi_{2z}}}^{2}}{ \left( 1+q \right) ^{4}}}
\right. \cr && \left.
+{\frac {167\,{\pi}^{2}q}{128\, \left( 1+q \right) ^{2}}}-{\frac {120\,
{q}^{6}+2744\,{q}^{5}+10049\,{q}^{4}+14820\,{q}^{3}+10049\,{q}^{2}+
2744\,q+120}{96\, \left( 1+q \right) ^{6}}} \right)
\left(\frac{{M}}{r}\right)^{3} \Biggr] \,,
\label{eq:rtoO}
\\
\frac{P_t}{M} &=&
\frac {q}{ \left( 1+q \right) ^{2}}\sqrt {\frac{M}{r}} 
\Biggl[ 1
+2\,{\frac {M}{r}}
+ \left( -\frac{3}{4}\,{\frac { \left( 3+4\,q \right) q{\chi_{1z}}}
{ \left( 1+q \right) ^{2}}}
-\frac{3}{4}\,{\frac { \left( 3\,q+4 \right) {\chi_{2z}}}
{ \left( 1+q \right) ^{2}}} \right)
\left(\frac{{M}}{r}\right)^{3/2}
\cr &&
+ \left( -\frac{3}{2}\,{\frac {{{\chi_{1x}}}^{2}{q}^{2}}
{ \left( 1+q \right) ^{2}}}
-3\,{\frac {{\chi_{1x}}\,{\chi_{2x}}\,q}{ \left( 1+q \right) ^{2}}}
+\frac{3}{4}\,{\frac {{{\chi_{1y}}}^{2}{q}^{2}}{ \left( 1+q \right) ^{2}}}
+\frac{3}{2}\,{\frac {{\chi_{1y}}\,{\chi_{2y}}\,q}{ \left( 1+q \right) ^{2}}}
+\frac{3}{4}\,{\frac {{{\chi_{1z}}}^{2}{q}^{2}}{ \left( 1+q \right) ^{2}}}
+\frac{3}{2}\,{\frac {{\chi_{1z}}\,{\chi_{2z}}\,q}{ \left( 1+q \right) ^{2}}}
\right. \cr && \left.
-\frac{3}{2}\,{\frac {{{\chi_{2x}}}^{2}}{ \left( 1+q \right) ^{2}}}
+\frac{3}{4}\,{\frac {{{\chi_{2y}}}^{2}}{ \left( 1+q \right) ^{2}}}
+\frac{3}{4}\,{\frac {{{\chi_{2z}}}^{2}}{ \left( 1+q \right) ^{2}}}
+\frac{1}{16}\,{\frac {42\,{q}^{2}+41\,q+42}{ \left( 1+q \right) ^{2}}} \right)
\left(\frac{{M}}{r}\right)^{2}
\cr &&
+ \left( -\frac{1}{16}\,{\frac {q \left( 72\,{q}^{3}+116\,{q}^{2}+60\,q+13 \right) 
{\chi_{1z}}}{ \left( 1+q \right) ^{4}}}
-\frac{1}{16}\,{\frac { \left( 13\,{q}^{3}+60\,{q}^{2}+116\,q+72 \right) {\chi_{2z}}}
{ \left( 1+q \right) ^{4}}} \right)
\left(\frac{{M}}{r}\right)^{5/2}
\cr &&
+ \left( -\frac{1}{16}\,{\frac {{q}^{2} \left( 80\,{q}^{2}-59 \right) 
{{\chi_{1x}}}^{2}}{ \left( 1+q \right) ^{4}}}
+\frac{1}{8}\,{\frac {q \left( 12\,{q}^{2}+35\,q+12 \right) 
{\chi_{2x}}\,{\chi_{1x}}}{ \left( 1+q \right) ^{4}}}
-\frac{1}{2}\,{\frac {{q}^{2} \left( {q}^{2}+10\,q+8 \right) {{\chi_{1y}}}^{2}}
{ \left( 1+q \right) ^{4}}}
\right. \cr && \left.
-\frac{1}{4}\,{\frac {q \left( 27\,{q}^{2}+58\,q+27 \right) {\chi_{2y}}\,{\chi_{1y}}}
{ \left( 1+q \right) ^{4}}}
+\frac{1}{32}\,{\frac {{q}^{2} \left( 128\,{q}^{2}+56\,q-27 \right) {{\chi_{1z}}}^{2}}
{ \left( 1+q \right) ^{4}}}
+\frac{1}{16}\,{\frac {q \left( 60\,{q}^{2}+133\,q+60 \right) {\chi_{2z}}\,{\chi_{1z}}}
{ \left( 1+q \right) ^{4}}}
\right. \cr && \left.
+\frac{1}{16}\,{\frac { \left( 59\,{q}^{2}-80 \right) {{\chi_{2x}}}^{2}}
{ \left( 1+q \right) ^{4}}}
-\frac{1}{2}\,{\frac { \left( 8\,{q}^{2}+10\,q+1 \right) {{\chi_{2y}}}^{2}}
{ \left( 1+q \right) ^{4}}}
-\frac{1}{32}\,{\frac { \left( 27\,{q}^{2}-56\,q-128 \right) {{\chi_{2z}}}^{2}}
{ \left( 1+q \right) ^{4}}}
\right. \cr && \left.
+{\frac {163\,{\pi}^{2}q}{128\, \left( 1+q \right) ^{2}}}
+\frac{1}{32}\,{\frac {120\,{q}^{4}-659\,{q}^{3}-1532\,{q}^{2}-659\,q+120}
{ \left( 1+q \right) ^{4}}} \right)
\left(\frac{{M}}{r}\right)^{3}
\Biggr] 
\,,
\\
\frac{M_{\rm ADM}}{M} &=&
1
-\frac{1}{2}\,{\frac {q}{ \left( 1+q \right) ^{2}}}\frac{{M}}{r}
+\frac{1}{8}\,{\frac {q \left( 7\,{q}^{2}+13\,q+7 \right)}
{ \left( 1+q \right) ^{4}}}
\left(\frac{{M}}{r}\right)^{2}
+ \left( -\frac{1}{4}\,{\frac {{q}^{2} \left( 3+4\,q \right) {\chi_{1z}}}
{ \left( 1+q \right) ^{4}}}
-\frac{1}{4}\,{\frac {q \left( 3\,q+4 \right) {\chi_{2z}}}
{ \left( 1+q \right) ^{4}}} \right)
\left(\frac{{M}}{r}\right)^{5/2}
\cr &&
+ \left( -\frac{1}{2}\,{\frac {{{\chi_{1x}}}^{2}{q}^{3}}
{ \left( 1+q \right) ^{4}}}
-{\frac {{\chi_{1x}}\,{\chi_{2x}}\,{q}^{2}}{ \left( 1+q \right) ^{4}}}
+\frac{1}{4}\,{\frac {{{\chi_{1y}}}^{2}{q}^{3}}{ \left( 1+q \right) ^{4}}}
+\frac{1}{2}\,{\frac {{\chi_{1y}}\,{\chi_{2y}}\,{q}^{2}}
{ \left( 1+q \right) ^{4}}}
+\frac{1}{4}\,{\frac {{{\chi_{1z}}}^{2}{q}^{3}}{ \left( 1+q \right) ^{4}}}
+\frac{1}{2}\,{\frac {{\chi_{1z}}\,{\chi_{2z}}\,{q}^{2}}
{ \left( 1+q \right) ^{4}}}
\right. \cr && \left.
-\frac{1}{2}\,{\frac {{{\chi_{2x}}}^{2}q}{ \left( 1+q \right) ^{4}}}
+\frac{1}{4}\,{\frac {{{\chi_{2y}}}^{2}q}{ \left( 1+q \right) ^{4}}}
+\frac{1}{4}\,{\frac {{{\chi_{2z}}}^{2}q}{ \left( 1+q \right) ^{4}}}
+\frac{1}{16}\,{\frac {q \left( 9\,{q}^{4}+16\,{q}^{3}+13\,{q}^{2}+16\,q+9 \right) }
{ \left( 1+q \right) ^{6}}} \right)
\left(\frac{{M}}{r}\right)^{3}
\cr &&
+ \left( -\frac{1}{16}\,{\frac {{q}^{2} \left( 32\,{q}^{3}+42\,{q}^{2}
+14\,q+1 \right) {\chi_{1z}}}{ \left( 1+q \right) ^{6}}}
-\frac{1}{16}\,{\frac {q \left( {q}^{3}+14\,{q}^{2}+42\,q+32 \right) {\chi_{2z}}}
{ \left( 1+q \right) ^{6}}} \right)
\left(\frac{{M}}{r}\right)^{7/2}
\cr &&
+ \left( -\frac{1}{16}\,{\frac { \left( 52\,{q}^{2}+12\,q-25 \right) {q}^{3}
{{\chi_{1x}}}^{2}}{ \left( 1+q \right) ^{6}}}
+{\frac {9\,{\chi_{1x}}\,{\chi_{2x}}\,{q}^{3}}
{8\, \left( 1+q \right) ^{6}}}
+\frac{1}{8}\,{\frac { \left( {q}^{2}-17\,q-15 \right) {q}^{3}{{\chi_{1y}}}^{2}}
{ \left( 1+q \right) ^{6}}}
\right. \cr && \left.
-\frac{3}{4}\,{\frac { \left( 4\,{q}^{2}+9\,q+4 \right) {q}^{2}
{\chi_{2y}}\,{\chi_{1y}}}{ \left( 1+q \right) ^{6}}}
+\frac{1}{16}\,{\frac { \left( 50\,{q}^{2}+38\,q+3 \right) {q}^{3}
{{\chi_{1z}}}^{2}}{ \left( 1+q \right) ^{6}}}
+\frac{3}{8}\,{\frac { \left( 10\,{q}^{2}+21\,q+10 \right) {q}^{2}
{\chi_{2z}}\,{\chi_{1z}}}{ \left( 1+q \right) ^{6}}}
\right. \cr && \left.
+\frac{1}{16}\,{\frac { \left( 25\,{q}^{2}-12\,q-52 \right) q{{\chi_{2x}}}^{2}}
{ \left( 1+q \right) ^{6}}}
-\frac{1}{8}\,{\frac { \left( 15\,{q}^{2}+17\,q-1 \right) q{{\chi_{2y}}}^{2}}
{ \left( 1+q \right) ^{6}}}
+\frac{1}{16}\,{\frac { \left( 3\,{q}^{2}+38\,q+50 \right) q{{\chi_{2z}}}^{2}}
{ \left( 1+q \right) ^{6}}}
\right. \cr && \left.
+{\frac {81\,{\pi}^{2}{q}^{2}}{128\, \left( 1+q \right) ^{4}}}
+{\frac {q \left( 537\,{q}^{6}-3497\,{q}^{5}-18707\,{q}^{4}
-29361\,{q}^{3}-18707\,{q}^{2}-3497\,q+537 \right) }
{384\, \left( 1+q \right) ^{8}}} \right)
\left(\frac{{M}}{r}\right)^{4}
\,,
\eea
\end{widetext}
where $P_{1y} = - P_{2 y} = P_t$.

In Appendix~\ref{app:ID_freq}, 
we present the ADM mass, orbital separation, and the tangential linear momentum
in terms of the orbital frequency $\Omega$.

\subsubsection{From the center-of-mass vector}

In the above analysis, we used the center-of-mass Hamiltonian, i.e.,
the center of mass was located at the origin and the vector $\vec r$
is the displacement $\vec r = \vec x_1 - \vec x_2$.
To obtain the positions of each BH, we need to know the position of
the BHs relative to the center of mass.
To do this, we use expressions for the center-of-mass (given BH
positions $\vec x_1$ and $\vec x_2$) based
on the nonspinning terms in Ref.~\cite{Damour:2000kk},
the spin-orbit terms in Ref.~\cite{Damour:2007nc},
the $S_1$-$S_2$ terms in Ref.~\cite{Steinhoff:2008zr},
and the spin-squared ($S^2$) terms in Ref.~\cite{Hergt:2008jn}.

Then, given that we want $\vec x_1 - \vec x_2$ to lie along the
$x$-axis, and the center of mass to lie on the origin, we solve
for $\vec x_1$ and $\vec x_2$.
The resulting expressions for the BH positions are given by
\begin{widetext}
\bea
\frac{x_1}{M} &=&
{\frac {1}{1+q}}\frac{r}{M}
-\frac{1}{2}\,{\frac {q \left( q-1 \right)}{ \left( 1+q \right) ^{3}}}
+\frac{1}{2}\,{\frac {\left( q-1 \right)  \left( 1+q \right) }{q}}
\frac{r {P_t}^{2}}{M^3}
+ \left( -\frac{1}{2}\,{\frac {q {\chi_{1z}}}{1+q}}
+\frac{1}{2}\,{\frac {{\chi_{2z}}}{1+q}} \right) \frac{P_t}{M}
\cr &&
+\frac{1}{4}\,{\frac {q\left( q-1 \right)  \left( {q}^{2}+1 \right) }
{\left( 1+q \right) ^{5} }}
\frac{M}{r}
-\frac{1}{4}\,{\frac { \left( q-1 \right)  \left( 5\,{q}^{2}+8\,q+5 \right)}
{q\left( 1+q \right) }}
\frac{ {P_t}^{2}}{M^2}
-\frac{1}{8}\,{\frac { \left( q-1 \right)  \left( 1+q \right) ^{5}}
{{q}^{3}}}
\frac{r{P_t}^{4}}{M^5}
\cr &&
+ \biggl[ -\frac{1}{2}\,{\frac {{q}^{2}\left( 5\,q-1 \right) {\chi_{1z}}}
{\left( 1+q \right) ^{3}}}
-\frac{1}{2}\,{\frac {\left( q-5 \right) {\chi_{2z}}}
{\left( 1+q \right) ^{3}}} \biggr]
\frac{P_t}{r}
\cr &&
+\frac{1}{16}\,{\frac { \left( q-1 \right)  \left( {q}^{2}+1 \right)  
\left( 1+q \right) ^{7}}{{q}^{5}}}
\frac{r{P_t}^{6}}{M^7}
+\frac{9}{16}\,{\frac { \left( q-1 \right)  \left( {q}^{2}+q+1 \right)  
\left( 1+q \right) ^{3}}{{q}^{3}}}
\frac{{P_t}^{4}}{M^4}
\cr &&
+\frac{1}{16}\,{\frac { \left( q-1 \right)  \left( 30\,{q}^{4}+125\,{q}^{3}
+198\,{q}^{2}+125\,q+30 \right)}{q \left( 1+q \right) ^{3}}}
\frac{ {P_t}^{2}}{Mr}
\cr &&
+\biggl[{\frac {{q}^{2} \left( q-1 \right) {\chi_{2x}}\,{\chi_{1x}}}
{\left( 1+q \right) ^{5}}}
-\frac{1}{2}\,{\frac {{q}^{2} \left( q-1 \right) {\chi_{2y}}\,{\chi_{1y}}}
{\left( 1+q \right) ^{5}}}
-\frac{1}{2}\,{\frac {{q}^{2} \left( q-1 \right) {\chi_{2z}}\,{\chi_{1z}}}
{ \left( 1+q \right) ^{5}}}
\cr &&
-\frac{1}{8}\,{\frac { \left( q-1 \right)  \left( {q}^{4}+3\,{q}^{3}+8\,{q}^{2}
+3\,q+1 \right) q}{ \left( 1+q \right) ^{7}}}\biggr]
\frac{M^2}{r^2}
\cr &&
+ \biggl[ \frac{1}{8}\,{\frac { \left( 1+q \right)  \left( 2\,q+1 \right) {\chi_{1z}}}
{q}}
-\frac{1}{8}\,{\frac { \left( q+2 \right)  \left( 1+q \right) {\chi_{2z}}}
{q}} \biggr]
\frac{{P_t}^{3}}{M^3} \,,
\cr
\frac{y_1}{M} = \frac{y_2}{M} &=&
\biggl[
\frac{1}{2}\,{\frac {{q}^{2}{\chi_{2y}}\,{\chi_{1x}}}
{ \left( 1+q \right) ^{4}}}
-\frac{1}{2}\,{\frac {{q}^{2}{\chi_{2x}}\,{\chi_{1y}}}
{ \left( 1+q \right) ^{4}}}\biggr]
\frac{M^2}{r^2}
\,,
\cr
\frac{z_1}{M} = \frac{z_2}{M} &=&
\left( \frac{1}{2}\,{\frac {q{\chi_{1x}}}{1+q}}
-\frac{1}{2}\,{\frac {{\chi_{2x}}}{1+q}} \right)
\frac{P_t}{M}
+\biggl[ \frac{1}{2}\,{\frac {{q}^{2}{\chi_{2z}}\,{\chi_{1x}}}
{ \left( 1+q \right) ^{4}}}
-\frac{1}{2}\,{\frac {{q}^{2}{\chi_{2x}}\,{\chi_{1z}}}
{ \left( 1+q \right) ^{4}}} \biggr]
\frac{M^2}{r^2}
\cr &&
+ \biggl[ -\frac{1}{8}\,{\frac { \left( 1+q \right)  
\left( 2\,q+1 \right) {\chi_{1x}}}{q}}
+\frac{1}{8}\,{\frac { \left( q+2 \right)  
\left( 1+q \right) {\chi_{2x}}}{q}} \biggr]
\frac{{P_t}^{3}}{M^3}
\cr &&
+ \biggl[ \frac{1}{4}\,{\frac {q \left( 6\,{q}^{2}+3\,q-5 \right) 
{\chi_{1x}}}{\left( 1+q \right) ^{3}}}
+\frac{1}{4}\,{\frac { \left( 5\,{q}^{2}-3\,q-6 \right) 
{\chi_{2x}}}{ \left( 1+q \right) ^{3}}} \biggr]
\frac{P_t}{r}
\,,
\eea
\end{widetext}
where $x_2 = r - x_1$.
Note that for the precessing case, the two BHs are displaced from the $x$
axis. However, as required $y_1-y_2 = z_1 - z_2 = 0$.

\subsubsection{From radiation reaction}

The radial momentum (velocity), which is driven by the emission of
gravitational waves and tidal heating drives the inspiral. To get
low-eccentricity orbital parameters, the corresponding radial momentum 
generally needs to be
included.

To obtain the radial momentum, we begin with the time derivative
of the ADM mass
\bea
M_{\rm ADM} = M + E_{\rm Orb} \,,
\eea
where $E_{\rm Orb}$ is the orbital energy for quasicircular orbits.
Both the orbital energy and the BH masses change during the inspiral
(the change in the latter is due to {\it tidal heating} effects).
Thus, we have
\bea
\frac{dM_{\rm ADM}}{dt} = \frac{dM}{dt} + \frac{dE_{\rm Orb}}{dt} \,.
\eea
The ADM mass loss produces the flux of gravitational wave energy
leaving the binary. Hence
\bea
-\frac{dE_{\rm GW}}{dt} - \frac{dM}{dt} = \frac{dE_{\rm Orb}}{dt} \,,
\eea
and since $dE_{\rm Orb}/dt = (dr/dt)(dE_{\rm Orb}/dr)$,
we have
\bea
\frac{dr}{dt} = - \left(\frac{dE_{\rm GW}}{dt}+\frac{dM}{dt}\right)
\left(\frac{dE_{\rm Orb}}{dr}\right)^{-1} \,.
\label{eq:balance}
\eea

For nonspinning and nonprecessing cases, we use the orbital energy 
and GW energy flux summarized in Ref.~\cite{Brown:2007jx},
and for the precessing case, we can use the formulas
summarized in Appendix A of Ref.~\cite{Ossokine:2015vda}.
Ref.~\cite{Brown:2007jx} uses the work of Alvi~\cite{Alvi:2001mx} to
calculate $dM/dt$. While higher-order correction to $dM/dt$ 
in the nonprecessing case are known (see, e.g.,
Ref.~\cite{Chatziioannou:2012gq}), additional correction for the
precessing case are not known. Thus we use the formula given in
Ref.~\cite{Brown:2007jx} for both the non-precessing and precessing cases. 

Given $E_{\rm Orb}$, $dE_{\rm GW}/dt$, and $dM/dt$, it is
straightforward to calculate $dr/dt$. However, what we need is a
formula for $P_r$, rather than $dr/dt$ itself.
To obtain this, we note that
\bea
\frac{dr}{dt} = \frac{\partial H}{\partial P_r} \,,
\eea
In the right hand side of the above equation,
we pick up only $O(P_r^0)$ and $O(P_r^1)$ terms
because $O(P_r^2)$ and higher order terms can be ignored in the PN approximation.
In practice, we have
\begin{widetext}
\bea
\frac{dr}{dt} &=& 
\Biggl[
{\frac { \left( 1+q \right) ^{2}}{q}}
-\frac{1}{2}\,{\frac {7\,{q}^{2}+15\,q+7}{q}}
\frac{M}{r}
+\frac{1}{8}\,{\frac { \left( 47\,{q}^{4}+229\,{q}^{3}+363\,{q}^{2}+229\,q
+47 \right) }{q{r}^{2} \left( 1+q \right) }}
\left(\frac{M}{r}\right)^{2}
\cr && \quad
+\left(\frac{1}{4}\,{\frac { \left( 12\,{q}^{2}+11\,q+4 \right) {\chi_{1z}}}
{\left( 1+q \right) }}
+\frac{1}{4}\,{\frac { \left( 4\,{q}^{2}+11\,q+12 \right) {\chi_{2z}}}
{\left( 1+q \right) q}}
\right)
\left(\frac{M}{r}\right)^{5/2}
\cr && \quad
+\left( -\frac{1}{16}\,{{\pi}^{2}}
-\frac{1}{48}\,{\frac { \left( 363\,{q}^{6}+2608\,{q}^{5}+7324\,{q}^{4}
+10161\,{q}^{3}+7324\,{q}^{2}+2608\,q+363 \right) }{q 
\left( 1+q \right) ^{4}}} \right.
\cr && \quad
+ \frac{1}{4}\,{\frac { \left( 18\,{q}^{2}+6\,q+5 \right) q{{\chi_{1x}}}^{2}}
{ \left( 1+q \right) ^{2}}}
+{\frac { \left( 3\,{q}^{2}-q+3 \right) {\chi_{2x}}\,{\chi_{1x}}}
{ \left( 1+q \right) ^{2}}}
-\frac{3}{4}\,{\frac { \left( 3\,{q}^{2}+q+1 \right) q{{\chi_{1y}}}^{2}}
{ \left( 1+q \right) ^{2}}}
\cr && \quad
-\frac{1}{2}\,{\frac { \left( 3\,{q}^{2}-2\,q+3 \right) 
{\chi_{2y}}\,{\chi_{1y}}}{ \left( 1+q \right) ^{2}}}
-\frac{3}{4}\,{\frac { \left( 3\,{q}^{2}+q+1 \right) q{{\chi_{1z}}}^{2}}
{ \left( 1+q \right) ^{2}}}
-\frac{1}{2}\,{\frac { \left( 3\,{q}^{2}-2\,q+3 \right) 
{\chi_{2z}}\,{\chi_{1z}}}{ \left( 1+q \right) ^{2}}}
\cr && \quad
\left.
+\frac{1}{4}\,{\frac { \left( 5\,{q}^{2}+6\,q+18 \right) {{\chi_{2x}}}^{2}}
{q \left( 1+q \right) ^{2}}}
-\frac{3}{4}\,{\frac { \left( {q}^{2}+q+3 \right) {{\chi_{2y}}}^{2}}
{q \left( 1+q \right) ^{2}}}
-\frac{3}{4}\,{\frac { \left( {q}^{2}+q+3 \right) {{\chi_{2z}}}^{2}}
{q \left( 1+q \right) ^{2}}} \right)
\left(\frac{M}{r}\right)^{3} \Biggr] \,
\frac{P_r}{M}
\cr &&
+ \Biggl[ \frac{1}{2}\,{\frac {{q}^{2}{\chi_{1x}}{\chi_{1y}}}
{ \left( 1+q \right) ^{4}}}
-\frac{1}{4}\,{\frac {{q}^{2}{\chi_{2y}}{\chi_{1x}}}
{ \left( 1+q \right) ^{4}}}
-\frac{1}{4}\,{\frac {{q}^{2}{\chi_{2x}}\,{\chi_{1y}}}
{ \left( 1+q \right) ^{4}}}
+\frac{1}{2}\,{\frac {{q}^{2}{\chi_{2y}}\,{\chi_{2x}}}
{ \left( 1+q \right) ^{4}}} 
\Biggr]
\left(\frac{M}{r}\right)^{7/2}
\,,
\eea
\end{widetext}
and we may solve this equation with respect to $P_r$.
This treatment is basically the same as the one presented
in Ref.~\cite{Buonanno:2005xu} if we ignore the $O(P_r^0)$ term.
Note that we have ignored effects due to the change in mass of the two
BHs, i.e., $dM/dt$. This is because the effects of $dM/dt$ are
comparable in magnitude to the 4PN corrections to  $dE_{\rm GW}/dt$. A
complete self-consistent formulation of these 4PN terms has not
not been completely worked out yet, and is thus also ignored here.

\subsection{Orbital Evolution}\label{sec:Evol}

Given the extent of some of the PN expressions used in the
integration of the equations of motion for the orbital evolution,
we do not provide them explicitly here,
but give a detailed set of references
where these expressions can be found.

Our original implementation of the 3PN orbital equations of motion in
the ADM-TT gauge 
(that for the sake of simplicity we will label this as ``3PNevolution'')
was first described in Ref.~\cite{Campanelli:2008nk}.
That code was based on the formulation developed in Ref.~\cite{Buonanno:2005xu}
in which the Hamiltonian has the form
\bea
H &=& H_{\rm O,Newt} + H_{\rm O,1PN} + H_{\rm O,2PN} + H_{\rm O,3PN}
\cr && + H_{\rm SO,1.5PN} + H_{\rm S_1S_2,2PN} + H_{\rm S^2,2PN} \,.
\eea
The equations of motion were obtained from  Eqs.~(2.23)-(2.25) and
(3.1) in Ref.~\cite{Buonanno:2005xu}, and the
radiation reaction force from Eq.~(3.27) there, as well.
In addition, we added higher-order PN terms derived in 
Refs.~\cite{Damour:2007nc,Steinhoff:2007mb,Steinhoff:2008ji}
to the above Hamiltonian, and
higher-order corrections to the radiation reaction
derived in Refs.~\cite{Arun:2008kb,Blanchet:2011zv}.
The Hamiltonian is then given by Eq.~\eqref{eq:3PNH}.

The new code that, for the sake of simplicity,
we will call the ``4PNevolution'' code includes,
in addition to the above expressions, higher-order terms 
we list below.
\begin{enumerate}[(i)]
\item
   New higher-order radiation reaction force terms from
   Ref.~\cite{Ossokine:2015vda}.
   We removed spin terms in Eq. (3.27) of Ref.~\cite{Buonanno:2005xu}
   because the orbital averaging of this force becomes zero.
   In the EOB approach, the same treatment has been done
   (see, e.g., Ref.~\cite{Pan:2009wj}).
   \item Spin $S^3$ and $S^4$ terms in the Hamiltonian~\cite{Levi:2014gsa}.
   \item 4PN nonspinning local term from Eq.~(5.13)
   in Ref.~\cite{Damour:2014jta}, as well as nonlocal terms 
   from Eq.~(7.9) (or (7.12a)) of Ref.~\cite{Damour:2015isa}.
   Note that we assume
   ``quasicircular'' orbits for this nonlocal term.
   The consistent result for the 4PN nonspinning term
   was recently derived in Ref.~\cite{Bernard:2016wrg}.
   \item 3.5PN spin-orbit coupling (NNLO SO) from
   Eq.~(140a) of Ref.~\cite{Hartung:2013dza} which is in
   in the center-of-mass frame.
   This has been confirmed in Ref.~\cite{Levi:2015uxa}.
   \item 4PN S1S2 coupling terms (NNLO S1S2) from
   Eq.~(140b) of Ref.~\cite{Hartung:2013dza}.
   This has been confirmed by Ref.~\cite{Levi:2014sba}
   (note that there was a typo in Eq.~(140b)).
\end{enumerate}
We note that we do not include any corrections for:
\begin{enumerate}[(i)]
   \item The change in mass of the BHs during the evolution.
   \item 4PN spin-squared (NNLO $S^2$) in the EFT gauge. It was 
   derived in Ref.~\cite{Levi:2016ofk}, but it has not yet
   been confirmed in the ADM-TT gauge.
\end{enumerate}

\section{Full Numerical Evolutions}\label{sec:FN}

\subsection{Methods}\label{sec:Methods}

Due to large computational expense of numerical relativity simulations
of merging BHBs, 
in order to make systematic studies and build a data bank of
full numerical simulations, it is crucial to develop efficient 
numerical algorithms. 
To this end, we evolve the following BHB data sets using the {\sc
LazEv}~\cite{Zlochower:2005bj} implementation of the moving puncture
approach~\cite{Campanelli:2005dd,Baker:2005vv} with the conformal
function $W=\sqrt{\chi}=\exp(-2\phi)$ suggested by
Ref.~\cite{Marronetti:2007wz}.  For the runs presented here, we use
centered, sixth-order finite differencing in
space~\cite{Lousto:2007rj} and a fourth-order Runge Kutta time
integrator. 
This sixth-order spatial finite differencing speeds up the code by
a factor of $4/3$ compared to an eighth-order implementation
(mostly, this is due to the reduction in the number of
ghostzones). 
We also used a Courant factor (CFL) of $1/3$ instead of the
previous CFL of $1/4$~\cite{Zlochower:2012fk},
gaining another speedup factor of 4/3. 
We verified that, for the runs presented here, increasing the CFL and
reducing the finite difference order still lead to acceptable
conservation of the BH masses and spins during the evolution,
as well as an acceptable gravitational wave phase error during
the entire simulation (below $10^{-5}$). 

This plus the use of the new
\xsede supercomputer {\it Comet} at 
SDSC~\footnote{https://portal.xsede.org/sdsc-comet} led to typical
evolution speeds of $250M/{\rm day}$ on 16 nodes. Note that our
previous~\cite{Lousto:2013oza, Lousto:2015uwa}
comparable simulations averaged $\sim100M/{\rm day}$.

Our code uses the {\sc EinsteinToolkit}~\cite{Loffler:2011ay,
einsteintoolkit} / {\sc Cactus}~\cite{cactus_web} /
{\sc Carpet}~\cite{Schnetter-etal-03b}
infrastructure.  The {\sc
Carpet} mesh refinement driver provides a
``moving boxes'' style of mesh refinement. In this approach, refined
grids of fixed size are arranged about the coordinate centers of both
holes.  The {\sc Carpet} code then moves these fine grids about the
computational domain by following the trajectories of the two BHs.

We use {\sc AHFinderDirect}~\cite{Thornburg2003:AH-finding} to locate
apparent horizons.  We measure the magnitude of the horizon spin using
the {\it isolated horizon} (IH) algorithm detailed in
Ref.~\cite{Dreyer02a} and as implemented in Ref.~\cite{Campanelli:2006fy}.
Note that once we have the horizon spin, we can calculate the horizon
mass via the Christodoulou formula 
${m_H} = \sqrt{m_{\rm irr}^2 + S_H^2/(4 m_{\rm irr}^2)}\,,$
where $m_{\rm irr} = \sqrt{A/(16 \pi)}$, $A$ is the surface area of
the horizon, and $S_H$ is the spin angular momentum of the BH (in
units of $M^2$).  

We use the Antenna code~\cite{Campanelli:2005ia} to calculate the
gravitational waveform via the Weyl scalar $\psi_4$. Here, we
decompose $\psi_4$ into $(\ell, m)$ modes. For the present work,
we analyze the $(\ell=2, m=\pm2)$ modes, in particular, because of the
relatively simple way the eccentricity of (non-precessing) binaries
can be extracted from them.

\subsection{Initial Orbital Parameters}\label{sec:ID}

To compute the numerical initial data, we use the puncture
approach~\cite{Brandt97b} along with the {\sc
TwoPunctures}~\cite{Ansorg:2004ds} thorn.  In this approach the
3-metric on the initial slice has the form $\gamma_{a b} = (\psi_{BL}
+ u)^4 \delta_{a b}$, where $\psi_{BL}$ is the Brill-Lindquist
conformal factor, $\delta_{ab}$ is the Euclidean metric, and $u$ is
(at least) $C^2$ on the punctures.  The Brill-Lindquist conformal
factor is given by $ \psi_{BL} = 1 + \sum_{i=1}^n m_{i}^p / (2 |\vec r
- \vec r_i|), $ where $n$ is the total number of `punctures',
$m_{i}^p$ is the mass parameter of puncture $i$ ($m_{i}^p$ is {\em
not} the horizon mass associated with puncture $i$), and $\vec r_i$ is
the coordinate location of puncture $i$. For the initial (conformal)
extrinsic
curvature we take the analytic form $\hat{K}_{ij}^{BY}$ given by
Bowen and York~\cite{Bowen80}. In the puncture formalism, there are
15 non-trivial
free parameters. These are the initial coordinate separation of the
two BHs, the three components of the linear momentum and spin of each
BH, and finally, the mass parameter of each BH.
The momentum, spin and separation parameters are obtained directly
from the various PN approximations described above. The mass
parameters, however, have to be set by demanding that the total ADM
mass matches the PN prediction and the mass ratio matches the desired
value.  

In this work, we evolve eight sets of simulations of
spinning, nonprecessing BHBs spanning a range of
mass ratios $1/3\leq q\leq1$, including nonspinning cases.
For each configuration,
we consider five different approximations to generate low-eccentricity
data. These are: 
\begin{itemize}
  \item[{\bf id0}] Quasicircular data using an incomplete 3PN
    Hamiltonian.

  \item[{\bf id1}] Quasicircular data with radiation reaction-driven
    radial momentum using  all 3PN terms.

  \item[{\bf id2}] Quasicircular data with no radial momentum using
    the full 3PN Hamiltonian.

  \item[{\bf id3}] Inspiral parameters from a 3PN evolution from large
    separations.

  \item[{\bf id4}] Inspiral parameters from a 4PN evolution from large
    separations.
\end{itemize}
The initial data parameters for the runs presented here are given in
Table~\ref{tab:ID} (note that we consider 5
momentum
variations of the same 8 basic configurations). In this paper, we
chose to provide initial data by specifying an initial orbital
separation rather than an initial orbital frequency (this choice is
convenient when obtaining the  parameters based from PN inspiral evolutions),
although the
formalism allows for specifying the initial orbital frequency instead.

\begin{table*}
  \caption{
Initial data parameters for the quasi-circular
configurations with a smaller mass black hole (labeled 1),
and a larger mass spinning black hole (labeled 2). 
The 40 configurations are split into 8 families (labeled \#1 -- \#8) of
fixed mass ratio, spins, and initial separations. Within each family,
5 different choices for the QC momentum parameters are used (id0 -- id4). For a given family, 
the parameters that remain fixed are given for id0 only.
The mass ratio $q = m_{H1}/m_{H2}$
is given in terms of the Christodoulou masses of each BH.
The parameters $m^p_{1,2}$ are chosen such that $m_{H1}$ and 
$m_{H2}$ agree with the PN parameters $m_{1,2}$ to at least 5
significant digits (this also means that $m_{H1}/M + m_{H2}/M = 1 + {\cal
O}(10^{-5})$).
The
dimensionless spins $\chi_1$ and $\chi_2$ were obtained using the IH
formalism. 
The orbital frequency $\Omega{\rm orb.}$ was measured directly from
the numerical orbital trajectory at $t=200M$.
The initial puncture locations are
$\vec r_1 = (x_1,0,0)$ and $\vec r_2 = (x_2,0,0)$
with mass parameters
$m^p/M$,   momentum $\vec P/M = \mp (P_r,
-P_t, 0)$, spins $\vec S_i = (0, 0, S_i)$.
The last column gives the ADM mass of each
configuration. Note that in the table $P_r$ is multiplied by $10^3$
and $P_t$ by 10.
}\label{tab:ID}
\begin{ruledtabular}
\begin{tabular}{lccccccccccccc}
  Run ID  & $q$ & $\chi_1$ & $\chi_2$ & $M\, \Omega_{\rm orb.}$ & $x_1/M$ & $x_2/M$  &
  $S_1/M^2$ & $S_2/M^2$ &  $m^p_1/M$ & $m^p_2/M$ & $10^3\cdot P_r/M$ &
  $10\cdot P_t/M$ & $M_{\rm ADM}$\\
\hline
\#1id0 & 0.333 & 0.8000 & -0.5000 & 0.0275 & -8.2500 & 2.7500 & 0.0500
& -0.2812 & 0.1491 & 0.6580 & 0.0000 & 0.6841 & 0.993076\\
\#1id1 & $\cdots$ & $\cdots$ & $\cdots$ & 0.0258& $\cdots$ & $\cdots$
& $\cdots$ & $\cdots$ & 0.1491 & 0.6580 & 0.4448 & 0.6880 & 0.993182\\
\#1id2 & $\cdots$ & $\cdots$ & $\cdots$ & 0.0259& $\cdots$ & $\cdots$
& $\cdots$ & $\cdots$ & 0.1491 & 0.6580 & 0.0000 & 0.6880 & 0.993182\\
\#1id3 & $\cdots$ & $\cdots$ & $\cdots$ & 0.0259& $\cdots$ & $\cdots$
& $\cdots$ & $\cdots$ & 0.1491 & 0.6580 & 0.4490 & 0.6879 & 0.993179\\
\#1id4 & $\cdots$ & $\cdots$ & $\cdots$ & 0.0257& $\cdots$ & $\cdots$
& $\cdots$ & $\cdots$ & 0.1491 & 0.6580 & 0.4551 & 0.6882 & 0.993188\\
\\
\#2id0 & 0.333 & -0.8000 & 0.5000 & 0.0277 & -7.8750 & 2.6250 &
-0.0500 & 0.2812 & 0.1489 & 0.6576 & 0.0000 & 0.6869 & 0.992637\\
\#2id1 & $\cdots$ & $\cdots$ & $\cdots$ & 0.0270& $\cdots$ & $\cdots$
& $\cdots$ & $\cdots$ & 0.1489 & 0.6576 & 0.4467 & 0.6888 & 0.992690\\
\#2id2 & $\cdots$ & $\cdots$ & $\cdots$ & 0.0272& $\cdots$ & $\cdots$
& $\cdots$ & $\cdots$ & 0.1489 & 0.6576 & 0.0000 & 0.6888 & 0.992689\\
\#2id3 & $\cdots$ & $\cdots$ & $\cdots$ & 0.0268& $\cdots$ & $\cdots$
& $\cdots$ & $\cdots$ & 0.1489 & 0.6576 & 0.4521 & 0.6893 & 0.992705\\
\#2id4 & $\cdots$ & $\cdots$ & $\cdots$ & 0.0254& $\cdots$ & $\cdots$
& $\cdots$ & $\cdots$ & 0.1489 & 0.6576 & 0.4831 & 0.6933 & 0.992818\\
\\
\#3id0 & 0.250 & 0.8000 & -0.8000 & 0.0316 & -8.4880 & 2.1220 & 0.0320
& -0.5120 & 0.1186 & 0.4928 & 0.0000 & 0.6114 & 0.994076\\
\#3id1 & $\cdots$ & $\cdots$ & $\cdots$ & 0.0277& $\cdots$ & $\cdots$
& $\cdots$ & $\cdots$ & 0.1186 & 0.4927 & 0.4196 & 0.6176 & 0.994249\\
\#3id2 & $\cdots$ & $\cdots$ & $\cdots$ & 0.0278& $\cdots$ & $\cdots$
& $\cdots$ & $\cdots$ & 0.1186 & 0.4927 & 0.0000 & 0.6176 & 0.994249\\
\#3id3 & $\cdots$ & $\cdots$ & $\cdots$ & 0.0283& $\cdots$ & $\cdots$
& $\cdots$ & $\cdots$ & 0.1186 & 0.4927 & 0.4302 & 0.6165 & 0.994220\\
\#3id4 & $\cdots$ & $\cdots$ & $\cdots$ & 0.0281& $\cdots$ & $\cdots$
& $\cdots$ & $\cdots$ & 0.1186 & 0.4927 & 0.4365 & 0.6169 & 0.994230\\
\\
\#4id0 & 1.000 & -0.8000 & -0.8000 & 0.0314 & -5.5000 & 5.5000 &
-0.2000 & -0.2000 & 0.3029 & 0.3029 & 0.0000 & 0.9411 & 0.991047\\
\#4id1 & $\cdots$ & $\cdots$ & $\cdots$ & 0.0269& $\cdots$ & $\cdots$
& $\cdots$ & $\cdots$ & 0.3029 & 0.3029 & 0.9437 & 0.9519 & 0.991351\\
\#4id2 & $\cdots$ & $\cdots$ & $\cdots$ & 0.0268& $\cdots$ & $\cdots$
& $\cdots$ & $\cdots$ & 0.3029 & 0.3029 & 0.0000 & 0.9519 & 0.991349\\
\#4id3 & $\cdots$ & $\cdots$ & $\cdots$ & 0.0299& $\cdots$ & $\cdots$
& $\cdots$ & $\cdots$ & 0.3029 & 0.3029 & 0.8941 & 0.9444 & 0.991140\\
\#4id4 & $\cdots$ & $\cdots$ & $\cdots$ & 0.0256& $\cdots$ & $\cdots$
& $\cdots$ & $\cdots$ & 0.3028 & 0.3028 & 1.0522 & 0.9554 & 0.991449\\
\\
\#5id0 & 0.750 & -0.8500 & 0.6375 & 0.0270 & -6.2857 & 4.7143 &
-0.1561 & 0.2082 & 0.2192 & 0.4479 & 0.0000 & 0.8779 & 0.990803\\
\#5id1 & $\cdots$ & $\cdots$ & $\cdots$ & 0.0255& $\cdots$ & $\cdots$
& $\cdots$ & $\cdots$ & 0.2192 & 0.4479 & 0.6879 & 0.8824 & 0.990926\\
\#5id2 & $\cdots$ & $\cdots$ & $\cdots$ & 0.0257& $\cdots$ & $\cdots$
& $\cdots$ & $\cdots$ & 0.2192 & 0.4479 & 0.0000 & 0.8824 & 0.990926\\
\#5id3 & $\cdots$ & $\cdots$ & $\cdots$ & 0.0258& $\cdots$ & $\cdots$
& $\cdots$ & $\cdots$ & 0.2192 & 0.4479 & 0.6788 & 0.8815 & 0.990902\\
\#5id4 & $\cdots$ & $\cdots$ & $\cdots$ & 0.0244& $\cdots$ & $\cdots$
& $\cdots$ & $\cdots$ & 0.2192 & 0.4479 & 0.7284 & 0.8863 & 0.991034\\
\\
\#6id0 & 0.700 & 0.0000 & 0.0000 & 0.0234 & -7.0588 & 4.9412 & 0.0000
& 0.0000 & 0.4002 & 0.5771 & 0.0000 & 0.8219 & 0.991345\\
\#6id1 & $\cdots$ & $\cdots$ & $\cdots$ & 0.0226& $\cdots$ & $\cdots$
& $\cdots$ & $\cdots$ & 0.4002 & 0.5771 & 0.5066 & 0.8246 & 0.991418\\
\#6id2 & $\cdots$ & $\cdots$ & $\cdots$ & 0.0226& $\cdots$ & $\cdots$
& $\cdots$ & $\cdots$ & 0.4002 & 0.5771 & 0.0000 & 0.8246 & 0.991417\\
\#6id3 & $\cdots$ & $\cdots$ & $\cdots$ & 0.0228& $\cdots$ & $\cdots$
& $\cdots$ & $\cdots$ & 0.4002 & 0.5771 & 0.5031 & 0.8241 & 0.991405\\
\#6id4 & $\cdots$ & $\cdots$ & $\cdots$ & 0.0221& $\cdots$ & $\cdots$
& $\cdots$ & $\cdots$ & 0.4002 & 0.5771 & 0.5227 & 0.8266 & 0.991471\\
\\
\#7id0 & 0.250 & 0.0000 & 0.0000 & 0.0277 & -8.4000 & 2.1000 & 0.0000
& 0.0000 & 0.1909 & 0.7920 & 0.0000 & 0.5959 & 0.993687\\
\#7id1 & $\cdots$ & $\cdots$ & $\cdots$ & 0.0268& $\cdots$ & $\cdots$
& $\cdots$ & $\cdots$ & 0.1909 & 0.7919 & 0.3541 & 0.5977 & 0.993737\\
\#7id2 & $\cdots$ & $\cdots$ & $\cdots$ & 0.0269& $\cdots$ & $\cdots$
& $\cdots$ & $\cdots$ & 0.1909 & 0.7919 & 0.0000 & 0.5977 & 0.993736\\
\#7id3 & $\cdots$ & $\cdots$ & $\cdots$ & 0.0265& $\cdots$ & $\cdots$
& $\cdots$ & $\cdots$ & 0.1909 & 0.7919 & 0.3626 & 0.5983 & 0.993754\\
\#7id4 & $\cdots$ & $\cdots$ & $\cdots$ & 0.0261& $\cdots$ & $\cdots$
& $\cdots$ & $\cdots$ & 0.1909 & 0.7919 & 0.3705 & 0.5994 & 0.993783\\
\\
\#8id0 & 0.333 & 0.8000 & 0.8000 & 0.0306 & -7.1250 & 2.3750 & 0.0500
& 0.4500 & 0.1483 & 0.4601 & 0.0000 & 0.7049 & 0.991784\\
\#8id1 & $\cdots$ & $\cdots$ & $\cdots$ & 0.0302& $\cdots$ & $\cdots$
& $\cdots$ & $\cdots$ & 0.1483 & 0.4601 & 0.5315 & 0.7047 & 0.991779\\
\#8id2 & $\cdots$ & $\cdots$ & $\cdots$ & 0.0306& $\cdots$ & $\cdots$
& $\cdots$ & $\cdots$ & 0.1483 & 0.4601 & 0.0000 & 0.7047 & 0.991778\\
\#8id3 & $\cdots$ & $\cdots$ & $\cdots$ & 0.0293& $\cdots$ & $\cdots$
& $\cdots$ & $\cdots$ & 0.1483 & 0.4601 & 0.5616 & 0.7082 & 0.991880\\
\#8id4 & $\cdots$ & $\cdots$ & $\cdots$ & 0.0297& $\cdots$ & $\cdots$
& $\cdots$ & $\cdots$ & 0.1483 & 0.4601 & 0.5333 & 0.7068 & 0.991841\\

\end{tabular}
\end{ruledtabular}
\end{table*}

\begin{figure}
  \includegraphics[angle=270,width=0.95\columnwidth]{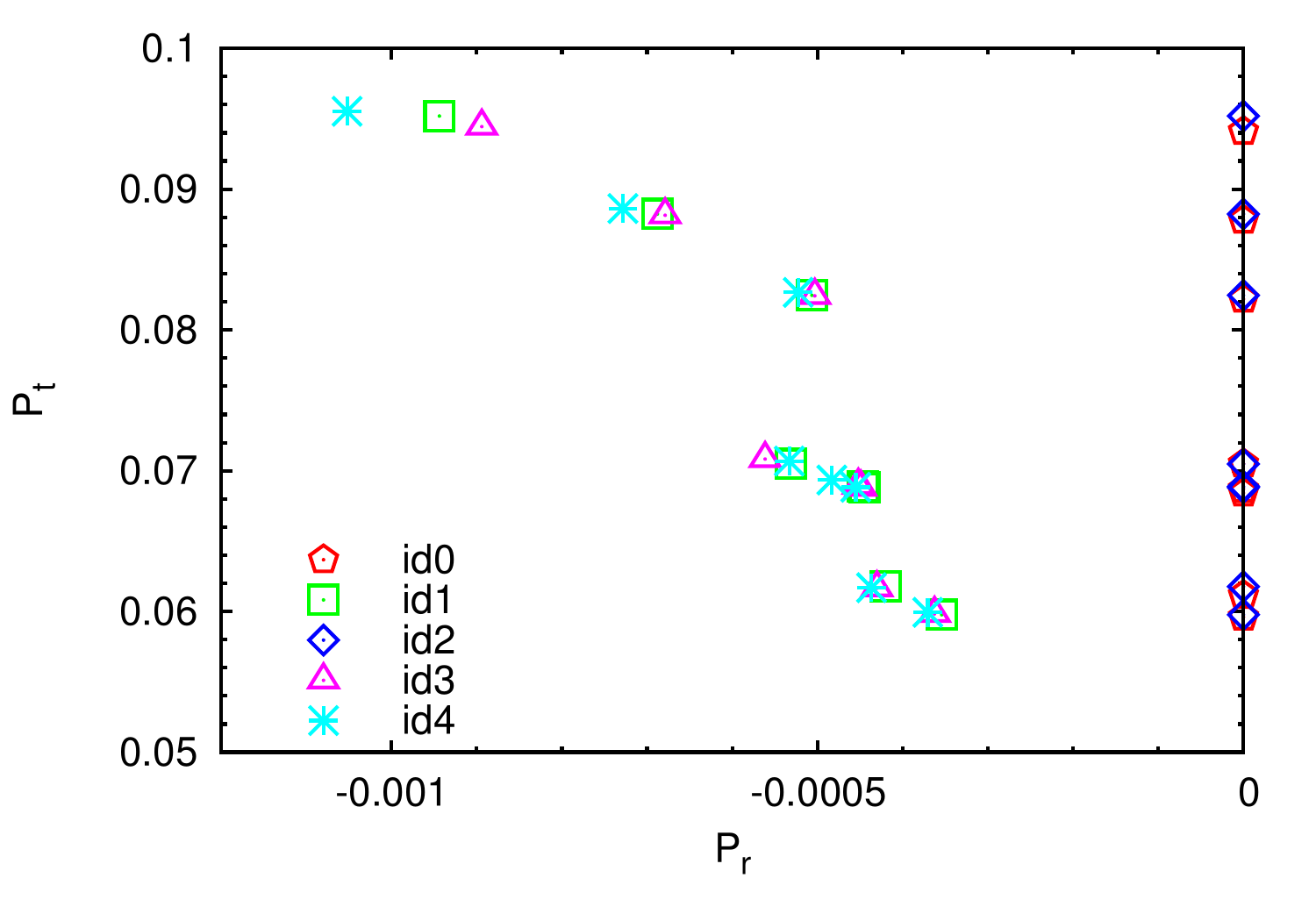}
  \caption{Initial parameters in the $(P_r, P_t)$ plane.
    Legends: id0 = QC incomplete 3PN.
id1 = QC complete 3PN, $P_r\not=0$. 
id2 = QC complete 3PN, $P_r=0$.
id3 = 3PNevolution.
id4 = 4PNevolution. See also Table~\ref{tab:ID}.
  \label{fig:params}}
\end{figure}

It is interesting to note how the various approximations change the
momentum parameters. In Fig.~\ref{fig:params}, we show the tangential
momentum versus radial momentum for all 40 configurations.

Finally, it is important to note that we are testing these variations
of choices of the initial data parameters in the context of
Brandt-Br\"ugmann puncture data. While  Brandt-Br\"ugmann data
are quite popular, the effects of the different
parameters choices on the eccentricities produced by other methods for
solving binary data, such as the extended conformal thin sandwich
methods used by the
Spectral Einstein Code (SpEC)~\cite{spec_web} for both conformally
flat~\cite{Cook:2004kt, Caudill:2006hw} and conformally
Kerr~\cite{Lovelace:2008tw} backgrounds, as well as on the
newly-developed puncture-based conformally Kerr
data~\cite{Ruchlin:2014zva}, are not determined here.

\subsection{Results}\label{sec:Results}

We measure eccentricity, $e$, using several orbital and waveform-based
methods. For our orbital based methods, we use
the time
variation of the proper distance of the coordinate line joining the
two BHs, which we refer to as the {\it simple proper distance}, or
SPD. More specifically, the SPD is the proper length of the part of
the coordinate ray joining the two centers that is outside both
apparent horizons.
\footnote{
We found in Ref.~\cite{Lousto:2013oza}, Figure 9, that the coordinate separation shows
substantial gauge effects  during the first few orbits. While the SPD
varies consistently on the orbital frequency, the coordinate
separation can vary at twice this frequency. We observe the same
effect in the runs in this paper, hence we choose to work with the SPD for eccentricity computations.}
In the Newtonian limit the separation vector ($\vec r$) between the
two BHs (i.e., Newtonian point particles) is given by
\begin{eqnarray}
  \vec r (t) &=& r(t)\left[\cos \Phi(t), \sin \Phi(t),
0\right] \,, \nonumber\\
  r(t) &=& A (1+ e\, \cos \Omega t) + {\cal O}(e^2) \,,\nonumber\\
\Phi(t) & =& \Omega t + 2 e \sin \Omega t + {\cal O}(e^2) \,.
\label{eq:NewtTraj}
\end{eqnarray}
We therefore define
two {\it distance} eccentricities as
\begin{equation}
  e_{D1} = {\rm Amp}\left(\frac{D(t) - D_{\rm sec}(t)}{D(t)}\right) \,,
  \label{ecc_N2}
\end{equation}
and
\begin{equation}
  e_{D2} = {\rm Amp}\left(\frac{D(t)^2 \ddot D(t)}{M_{\rm ADM}}\right) \,,
  \label{ecc_N1}
\end{equation}
where $D(t)$ is the numerical SPD, a dot
indicates a coordinate time derivative,
$D_{\rm sec}(t)$ is the non-oscillatory part of $D(t)$, and
${\rm Amp}$ indicates the amplitude of a sinusoidal function. In practice,
$D_{\rm sec}$ is constructed by fitting $D(t)$ to a low-order
polynomial in $\sqrt{t}$ plus a sinusoidal function. The polynomial
part of the fit is $D_{\rm sec}$. The difference $D(t)-D_{\rm sec}(t)$  is then dominated by
the sinusoidal part of $D(t)$. 
Both
$e_{D1}$ and $e_{D2}$ are only accurate to linear order in the
eccentricity for Newtonian orbits. Note that 
$e_{D1}$ was introduced
in Ref.~\cite{Pfeiffer:2007yz} and 
$e_{D2}$ was introduced in Ref.~\cite{Campanelli:2008nk}.

\begin{figure}
  \includegraphics[width=0.95\columnwidth]{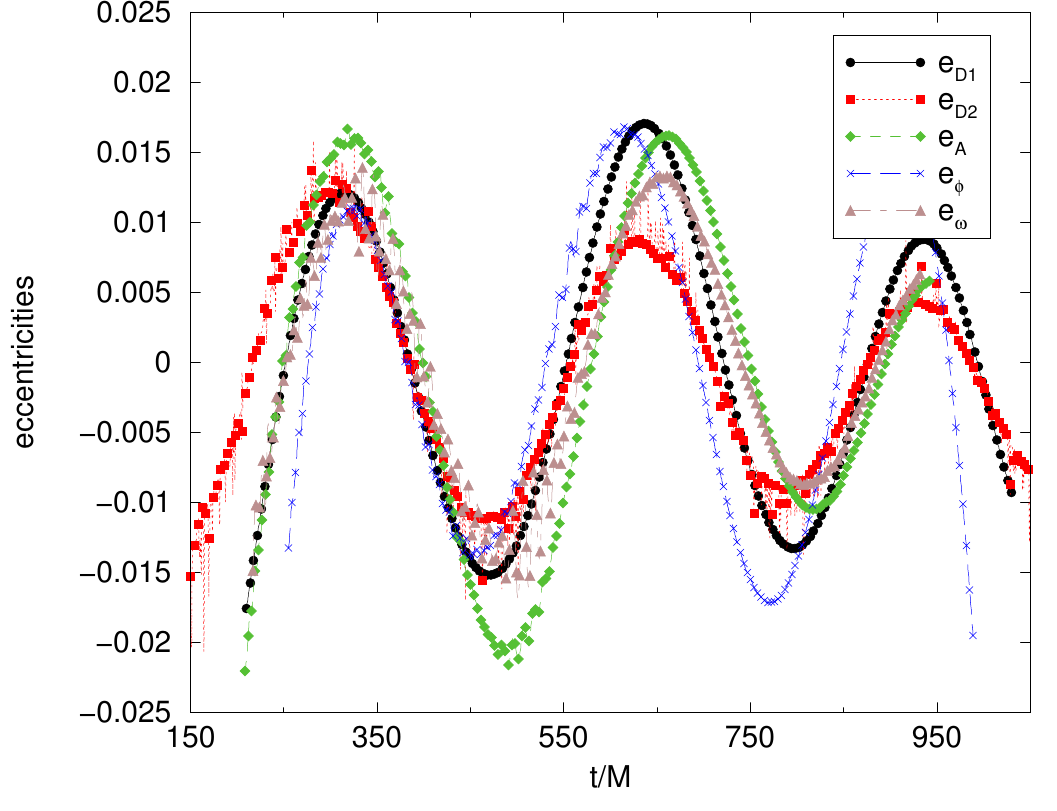}
  \caption{A comparison of the various measures of the eccentricity for
  run \#1, id0. The waveform measures ($e_A$, $e_\phi$, and
  $e_\omega$)
  have been translated by $t=100M$ to overlap with trajectory
  measures ($e_{D1}$ and
$e_{D2}$).}
  \label{fig:compare_measures}
\end{figure}

Following Ref.~\cite{Mroue:2010re},
we similarly define three eccentricities based on the amplitude,
frequency,  and
phase of the $(\ell=2,\,m=2)$ mode of the Weyl scalar,
$\psi_{4(22)}$. 
To find how the eccentricity affects the waveform, we start with the quadrapole formula (see, e.g.,
Ref.~\cite{Carroll:2004st})
\begin{equation}
  h^{\rm TT} = \frac{2}{R} \left.\frac{d^2 J_{ij}^{\rm TT}}{d t^2}\right|_{\rm ret} \bar m^i \bar m^j \,,
  \label{eq:hTT}
\end{equation}
where $J_{ij}^{\rm TT}$ is transverse and traceless part of the
quadrapole moment (evaluated at retarded time), the complex null-vector $m^j$ is given by
\begin{equation}
  m^j=\frac{\partial x^j}{\partial \theta} + \frac{i}{\sin
  \theta}\frac{\partial x^j}{\partial \phi} \,,
\end{equation} 
  $\theta$ and $\phi$ are the standard
spherical-polar coordinates, and $R$ is the distance from the binary
to the observer. To obtain $\psi_4$, we take the second
time derivative of the $h^{\rm TT}$. 
For two point particles with separation vector $\vec D(t)$ given by
the low-eccentricity Newtonian trajectory for two point particles,
i.e., Eq.~(\ref{eq:NewtTraj}),
$\psi_{4(22)}$ is
given by
\begin{equation}
\psi_{4(22)} = A_{22}(t) \exp [i \varphi_{22}(t)] \,,
\end{equation}
where
\begin{eqnarray}
  A_{22}(t)  &=& K
  \left(1 + \frac{39}{8} e \cos \Omega t\right) + {\cal O}(e^2) \,,\\
  \varphi_{22}(t) &=& -2 \Omega t - \frac{21}{4} e \sin \Omega t + {\cal
  O}(e^2) \,,\\
  \omega_{22} &=& -2 \Omega \left(1 +\frac{21}{8} e \cos \Omega
t\right) + {\cal
  O}(e^2) \,,
\end{eqnarray}
$K$ is an overall normalization factor (given by
$$K=\frac{64 \mu m^2}{R A^4} \sqrt{\frac{\pi}{5}}$$ in the Newtonian
limit, where
$m$ is the total mass of the binary,
$\mu$ is its reduced mass, $A$ is the average separation of the
binary, and $R$ is the distance from the binary to
the observer)
and $\omega_{22} = d\varphi_{22}/dt$.
Hence we define three waveform-based eccentricity measures
\begin{equation}
  e_A = \left(\frac{8}{39}\right){\rm Amp}\left(\frac{A_{22}(t) -
  A_{22 \rm sec}(t)}{A_{22}(t)}\right) \,,
  \label{ecc_eA}
\end{equation}
\begin{equation}
  e_\omega = \left(\frac{8}{21}\right){\rm Amp}\left(\frac{\omega_{22}(t) - \omega_{22 \rm sec}(t)}{
  \omega_{22}(t)}\right) \,,
\label{ecc_ew}
\end{equation}
and
\begin{equation}
  e_\phi =\left(\frac{4}{21}\right){\rm Amp}\left[ \varphi_{22}(t) - \varphi_{22 \rm
sec}(t)\right] \,.\label{ecc_ep}
\end{equation}
We choose to define
eccentricities using $\psi_4$ rather than $h^{\rm TT}$ because $\psi_4$ is
directly calculated from the simulation results whereas $h^{\rm TT}$ requires a
double time integral. We compare the performance of each of these
measures of the eccentricity in Fig.~\ref{fig:compare_measures}. 
Note
how all are roughly equivalent for run \#1. Note that $e_{\phi}$ and $e_{\omega}$
were introduced in Ref.~\cite{Mroue:2010re}, but with slightly different coefficients
than
presented here. Note that due to the need for fitting the secular
part, all eccentricity measures, except $e_{D2}$, can suffer from a
significant biases due to the form of the fitting function and chosen
interval for the fit. However, $e_{D2}$ can be calculated directly
from simulation data and, while with some more residual gauge dependence 
than $e_A$, $e_\phi$, and $e_\omega$, $e_{D2}$ provides a critical
sanity check for those other measures.

For reference, we also include eccentricity measures based on
$h^{\rm TT}$. The $(\ell =2, m=2)$ mode of $h^{TT}$ [see
Eq.~(\ref{eq:hTT})]
is given by
\begin{equation}
  h_{(22)} = B_{22}(t) e^{i \vartheta_{22}(t)},
\end{equation}
where
\begin{eqnarray}
  B_{22}(t) &=& \frac{16 m \mu}{A R}\sqrt{\frac{\pi}{5}}\left(1 + \frac{3}{2} e
\cos\Omega t\right) + {\cal O}(e^2),\\
\vartheta_{22}(t) &=& -2 \Omega t - 3 e \sin \Omega t + {\cal O}(e^2),\\
\varpi_{22}(t) &=& -2 \Omega \left( 1 +\frac{3}{2} e \cos \Omega t
\right) + {\cal O}(e^2),
\end{eqnarray}
and $\varpi_{22} = d\vartheta_{22}/dt$.
The corresponding eccentricity measures are
\begin{equation}
  e_B = \left(\frac{2}{3}\right){\rm Amp}\left(\frac{B_{22}(t) -
  B_{22 \rm sec}(t)}{B_{22}(t)}\right) \,,
  \label{ecc_eB_h}
\end{equation}
\begin{equation}
  e_\varpi = \left(\frac{2}{3}\right){\rm
  Amp}\left(\frac{\varpi_{22}(t) - \varpi_{22 \rm sec}(t)}{
  \varpi_{22}(t)}\right) \,,
\label{ecc_ew_h}
\end{equation}
and
\begin{equation}
  e_\vartheta =\left(\frac{1}{3}\right){\rm Amp}\left[
    \vartheta_{22}(t) - \vartheta_{22 \rm
sec}(t)\right] \,.\label{ecc_ep_h}
\end{equation}

\begin{figure}
  \includegraphics[angle=270,width=0.95\columnwidth]{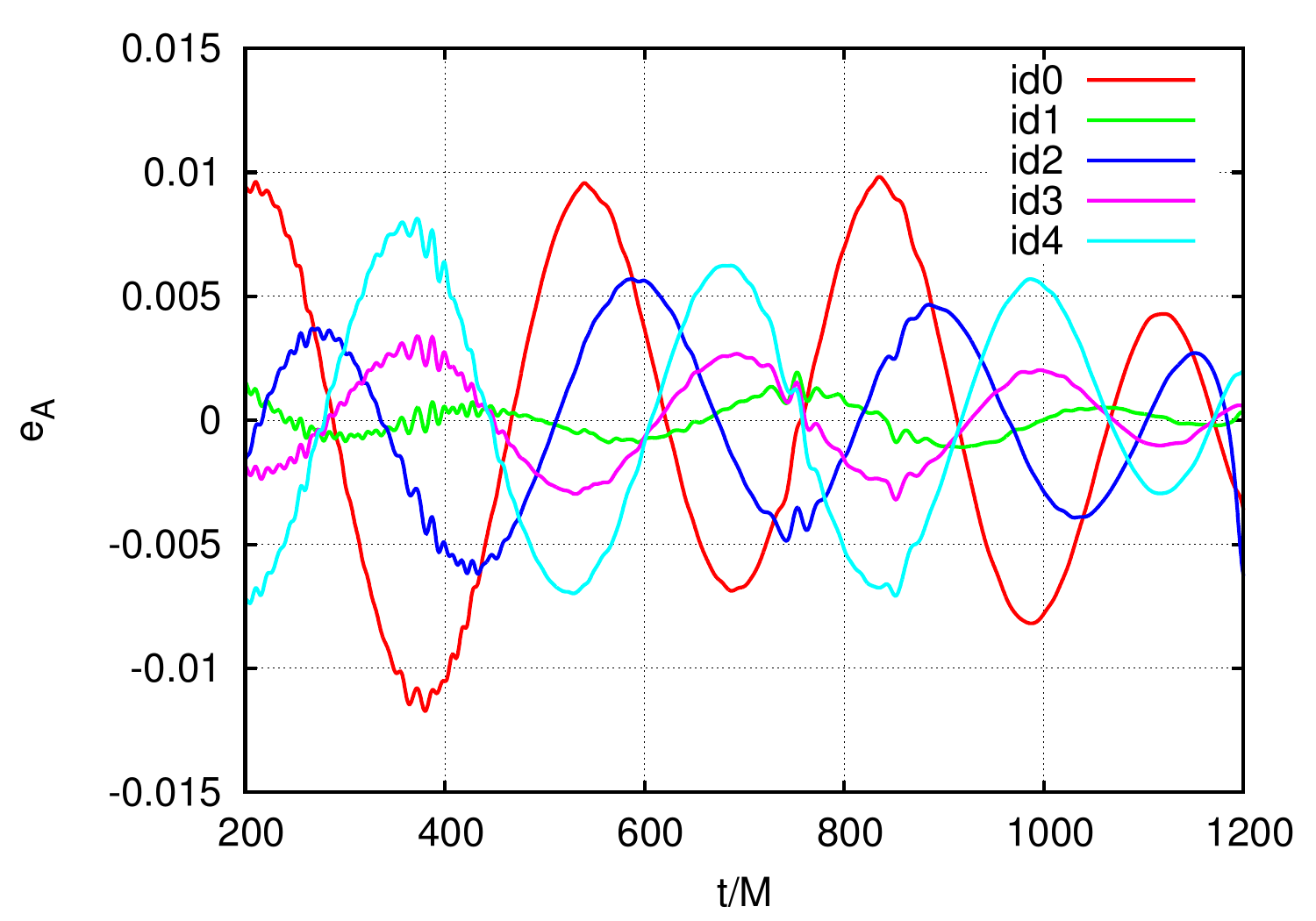}
  \caption{A comparison of the various initial data for run \#7
and the measures of the eccentricity from the waveform amplitude, Eq.~\eqref{ecc_eA}}
  \label{fig:run7e}
\end{figure}

The main result of this work is the comparison of the performance of
the 5 choices for low-eccentricity data. In most cases, id0 produced
the most eccentric data, and 
id1 produced the
best. Similarly, in most cases, id4 (4PN inspiral parameters) were higher
eccentricity than id3 (3PN inspiral parameters). We show the
sinusoidal dependence of $e_A$ for all five initial data
variations for run \#7 in Fig.~\ref{fig:run7e}. The eccentricities are
given by the amplitudes of these oscillations.
In Table~\ref{tab:edA},
we provide the eccentricities $e_{D2}$ and $e_{A}$
(representative of using trajectories and waveforms respectively)
for all
configurations. As both of the these are approximations, the two are
not supposed to agree exactly. We see about a $\lesssim20\%$ difference between
the two measures for most configurations. The largest relative differences
occur at low-eccentricities. This is due to waveform and trajectory
noise interfering with our ability to determine the secular and
oscillating parts of the amplitude of $\psi_4$ and the
second-time-derivative of $D$.

%
%
%
%

\begin{figure}
  \includegraphics[angle=270,width=0.95\columnwidth]{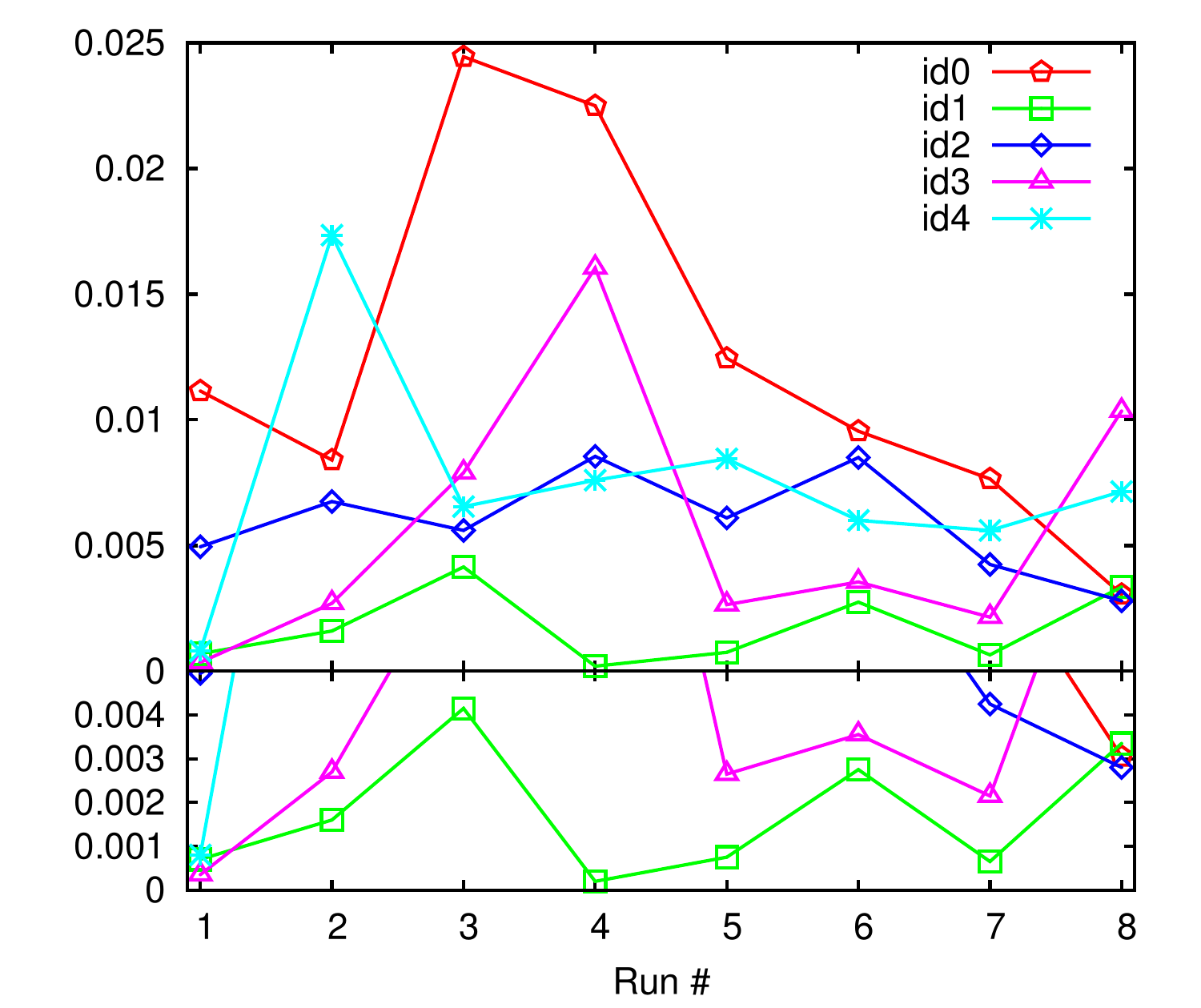}
  \caption{The eccentricity for each configuration as measured by the
    average of $e_{D2}$ and $e_A$. The bottom panel is a zoom in of the
    top one. Of the choices, id1 is consistently a top performer;
giving only a slightly larger eccentricity than id3 for one
configuration.} \label{fig:avg_ecc} \end{figure}

In Fig.~\ref{fig:avg_ecc}, we show the average of $e_{D2}$ and $e_{A}$
for all configurations. From the plot, it is clear that id1 is the
best overall choice for initial parameters. On average, the next
best choice is id3, which was proposed in Ref.~\cite{Husa:2007rh}.
Perhaps surprisingly, id4 is, on average, worse than id3. This
is remarkable because id4 is generated with higher-order PN terms than
id3 (both are based on PN inspirals).
This suggest that completing
all 3PN terms (id2) and including the radial momentum (id1) consistently
leads to the best results regarding reduction of eccentricities.
Notably, the evolution from large separations may lead to some
initial data with higher eccentricities (run \#4 for 3PN and
run \#2 for 4PN).

\begin{table*}
  \caption{The measured eccentricities using $e_{D2}$ and $e_{A}$ 
   for all 40 runs, as well as the relative difference between these
   two measures. Note that the relative differences are largest for
   small eccentricities due the waveform and trajectory noise
   dominating the small eccentricity effects. The eccentricities are given in the form
 $(e_{D2}, e_{A})$ [rel. difference]  in the table below.}\label{tab:edA}
\begin{ruledtabular}
\begin{tabular}{lccccc}
run\#  &id0    	&id1    	&id2         &id3   	    &id4\\
\hline
1 & (0.0120, 0.0103) [7\%]& (0.0007, 0.0007) [5\%]& (0.0051, 0.0048)
[3\%]& (0.0007, 0.0000) [100\%]& (0.0008, 0.0008) [1\%]\\
  2 & (0.0086, 0.0082) [2\%]& (0.0017, 0.0015) [7\%]& (0.0072, 0.0063)
[7\%]& (0.0028, 0.0026) [4\%]& (0.0200, 0.0147) [15\%]\\
  3 & (0.0227, 0.0262) [7\%]& (0.0032, 0.0051) [23\%]& (0.0060,
0.0052) [7\%]& (0.0062, 0.0096) [22\%]& (0.0051, 0.0080) [22\%]\\
  4 & (0.0250, 0.0200) [11\%]& (0.0004, 0.0000) [100\%]& (0.0081,
0.0090) [5\%]& (0.0170, 0.0151) [6\%]& (0.0090, 0.0062) [19\%]\\
  5 & (0.0133, 0.0116) [7\%]& (0.0006, 0.0009) [19\%]& (0.0069,
0.0053) [13\%]& (0.0026, 0.0027) [1\%]& (0.0093, 0.0076) [10\%]\\
  6 & (0.0100, 0.0091) [5\%]& (0.0035, 0.0020) [28\%]& (0.0086,
0.0084) [1\%]& (0.0036, 0.0035) [2\%]& (0.0062, 0.0058) [3\%]\\
  7 & (0.0072, 0.0081) [6\%]& (0.0006, 0.0007) [9\%]& (0.0042, 0.0043)
[1\%]& (0.0021, 0.0022) [3\%]& (0.0053, 0.0059) [6\%]\\
  8 & (0.0031, 0.0030) [2\%]& (0.0033, 0.0034) [2\%]& (0.0029, 0.0027)
[4\%]& (0.0100, 0.0107) [4\%]& (0.0069, 0.0074) [4\%]\\
\end{tabular}
\end{ruledtabular}
\end{table*}

The observation that our 4PN evolution does not give the lowest 
initial eccentricity may be a consequence of the poor convergence of the PN series.
Although one cannot make a strong statement concerning this for the
comparable mass case,
we note that the 3PN calculation has a larger region of validity
than 4PN for extreme mass ratio inspirals
(see, e.g., Refs.~\cite{Yunes:2008tw,Zhang:2011vha,Sago:2016xsp}).

\section{Conclusions and Discussion}\label{sec:Discussion}

In this paper, we measured the eccentricity resulting from fully
non-linear numerical evolutions of BHBs using orbital parameters
obtained from various post-Newtonian approximations.
We find that using the full 3PN Hamiltonian to generate quasicircular
orbital parameters and the addition of a radiation reaction-driven
inspiral momentum leads to significantly lower eccentricities than
the other methods we tried. This includes inspiral
parameters based on 3PN and 4PN evolutions from large separations.
Interestingly, we find that the (incomplete) 4PN inspiral parameters 
are sometimes more eccentric than the 3PN parameters.



These new quasicircular (plus radial) parameters 
can also serve as the
initial seed for iterative  methods for further reducing the
eccentricity (such as in 
Refs.~\cite{Pfeiffer:2007yz,Buonanno:2010yk,Purrer:2012wy}). Choosing
good starting parameters will reduce the number of iterations
required and thus reduce the overall computational cost considerably.

An interesting application of these measurements of eccentricity
from the waveforms, or more precisely, the analogs which directly use
the strain, i.e., $e_\varpi$, $e_\vartheta$, and $e_B$, is that they
can be used to give upper bounds to the binary's eccentricity
based solely on the observed waveform.  Our measures here do not
include precession effects, which can mimic true eccentricity, so
these measurements would provide an upper bound.

An open question for follow-up work will be to find similarly
performing initial parameters for highly-precessing binaries, as well
\cite{Buonanno:2010yk}. 
The quasicircular data given in Sec.~\ref{sec:QC}, 
and 3PN and 4PN evolutions described in Sec.~\ref{sec:Evol}
are applicable to precessing cases.

\acknowledgments 
The authors gratefully acknowledge the NSF for financial support from
Grants No.~PHY-1607520, No.~PHY-1707946 , No.~ACI-1550436 , No.~AST-1516150,
No.~ACI-1516125. Computational resources were provided by XSEDE
allocation TG-PHY060027N, and by NewHorizons and BlueSky Clusters at
Rochester Institute of Technology, which were supported by NSF grant
No.~PHY-0722703, No.~DMS-0820923, No.~AST-1028087, and No.~PHY-1229173.
HN is also supported by MEXT Grant-in-Aid for Scientific Research on
Innovative Areas, ``New Developments in Astrophysics Through
Multi-Messenger Observations of Gravitational Wave Sources'',
No.~24103006, and JSPS Grant-in-Aid for Scientific Research (C),
No.~16K05347.

\appendix

\section{Initial parameters in terms of the orbital frequency}\label{app:ID_freq}

The relationship $\Omega(r)$ in Eq.~\eqref{eq:rtoO}
can be inverted to solve for $r(\Omega)$,
and hence all quasicircular orbital parameters can be expressed as a
function of the orbital frequency.
In terms of the orbital frequency $\Omega$,
the ADM mass, orbital separation, and the tangential linear momentum
are given by
\begin{widetext}
\bea
\frac{M_{\rm ADM}}{M} &=&
1+\frac{q}{ \left( 1+q \right) ^{2}} \Biggl[
-\frac{1}{2}\, \left( M\Omega \right) ^{2/3}
+\frac{1}{24}
\,{\frac { \left( 9\,{q}^{2}+19\,q+9 \right) }
{ \left( 1+q \right) ^{2}}} \left( M\Omega \right) ^{4/3}
\cr &&
+
 \left( -\frac{1}{3}\,{\frac { \left( 3+4\,q \right) q{\chi_{1z}}}
{ \left( 1+q \right) ^{2}}}
-\frac{1}{3}\,{\frac { \left( 3\,q+4 \right) {\chi_{2z}}}
{ \left( 1+q \right) ^{2}}} \right)  \left( M\Omega \right) ^{5/3}
\cr &&
+ \left( -{\frac {{{\chi_{1x}}}^{2}{q}^{2}}{ \left( 1+q \right) ^{2}}}
-2\,{\frac {{\chi_{1x}}\,{\chi_{2x}}\,q}{ \left( 1+q \right) ^{2}}}
+\frac{1}{2}\,{\frac {{{\chi_{1y}}}^{2}{q}^{2}}{ \left( 1+q \right) ^{2}}}
+{\frac {{\chi_{1y}}\,{\chi_{2y}}\,q}{ \left( 1+q \right) ^{2}}}
+\frac{1}{2}\,{\frac {{{\chi_{1z}}}^{2}{q}^{2}}{ \left( 1+q \right) ^{2}}}
+{\frac {{\chi_{1z}}\,{\chi_{2z}}\,q}{ \left( 1+q \right) ^{2}}}
-{\frac {{{\chi_{2x}}}^{2}}{ \left( 1+q \right) ^{2}}}
\right. \cr && \quad \left.
+\frac{1}{2}\,{\frac {{{\chi_{2y}}}^{2}}{ \left( 1+q \right) ^{2}}}
+\frac{1}{2}\,{\frac {{{\chi_{2z}}}^{2}}{ \left( 1+q \right) ^{2}}}
+\frac{1}{48}\,{\frac {81\,{q}^{4}+267\,{q}^{3}+373\,{q}^{2}+267\,q+81}
{ \left( 1+q \right) ^{4}}} \right) \left({M}{\Omega}\right)^{2}
\cr &&
+ \left( -\frac{1}{18}\,{\frac {q \left( 72\,{q}^{3}+140\,{q}^{2}+96\,q+27 \right) 
{\chi_{1z}}}{ \left( 1+q \right) ^{4}}}
-\frac{1}{18}\,{\frac { \left( 27\,{q}^{3}+96\,{q}^{2}+140\,q+72 \right) {\chi_{2z}}}
{ \left( 1+q \right) ^{4}}} \right)  \left( M\Omega \right) ^{7/3}
\cr &&
+ \left( -{\frac {5\,{q}^{2} \left( 20\,{q}^{2}+4\,q-11 \right) 
{{\chi_{1x}}}^{2}}{24\, \left( 1+q \right) ^{4}}}
+\frac{5}{4}\,{\frac {{\chi_{1x}}\,{\chi_{2x}}\,{q}^{2}}{ \left( 1+q \right) ^{4}}}
-{\frac {5\,{q}^{2} \left( {q}^{2}+9\,q+7 \right) {{\chi_{1y}}}^{2}}
{12\, \left( 1+q \right) ^{4}}}
\right. \cr && \quad \left.
-\frac{6}{5}\,{\frac {q \left( 2\,q+3 \right)  \left( 3\,q+2 \right) 
{\chi_{2y}}\,{\chi_{1y}}}{ \left( 1+q \right) ^{4}}}+{\frac {5\,{q}^{2} 
\left( 13\,{q}^{2}-3\,q-9 \right) {{\chi_{1z}}}^{2}}{36\, \left( 1+q \right) ^{4}}}
+{\frac {5\,q \left( 3\,{q}^{2}+7\,q+3 \right) {\chi_{2z}}\,{\chi_{1z}}}
{18\, \left( 1+q \right) ^{4}}}
\right. \cr && \quad \left.
+{\frac { \left( 55\,{q}^{2}-20\,q-100 \right) 
{{\chi_{2x}}}^{2}}{24\, \left( 1+q \right) ^{4}}}
-{\frac { \left( 35\,{q}^{2}+45\,q+5 \right) {{\chi_{2y}}}^{2}}
{12\, \left( 1+q \right) ^{4}}}-{\frac { \left( 45\,{q}^{2}+15\,q-65 \right) 
{{\chi_{2z}}}^{2}}{36\, \left( 1+q \right) ^{4}}}
+{\frac {205\,{\pi}^{2}q}{192\, \left( 1+q \right) ^{2}}}
\right. \cr && \quad \left.
+{\frac {54675\,{q}^{6}+18045\,{q}^{5}-411525\,{q}^{4}-749755\,{q}^{3}
-411525\,{q}^{2}+18045\,q+54675}{10368\, \left( 1+q \right) ^{6}}} \right)
\left( M\Omega \right) ^{8/3} \Biggr] \,,
\label{eq:ADM_mass}
\\
\frac{r}{M}
&=& \left( M\Omega \right) ^{-2/3}
-\frac{1}{3}\,{\frac {3\,{q}^{2}+5\,q+3}{ \left( 1+q \right) ^{2}}}
+ \left( -\frac{1}{6}\,{\frac { \left( 3+4\,q \right) q{\chi_{1z}}}
{ \left( 1+q \right) ^{2}}}
-\frac{1}{6}\,{\frac { \left( 3\,q+4 \right) {\chi_{2z}}}
{ \left( 1+q \right) ^{2}}} \right) \left( M\Omega \right) ^{1/3}
\cr &&
+ \left( -{\frac {{{\chi_{1x}}}^{2}{q}^{2}}{ \left( 1+q \right) ^{2}}}
-2\,{\frac {{\chi_{1x}}\,{\chi_{2x}}\,q}{ \left( 1+q \right) ^{2}}}
+\frac{1}{2}\,{\frac {{{\chi_{1y}}}^{2}{q}^{2}}{ \left( 1+q \right) ^{2}}}
+{\frac {{\chi_{1y}}\,{\chi_{2y}}\,q}{ \left( 1+q \right) ^{2}}}
+\frac{1}{2}\,{\frac {{{\chi_{1z}}}^{2}{q}^{2}}{ \left( 1+q \right) ^{2}}}
+{\frac {{\chi_{1z}}\,{\chi_{2z}}\,q}{ \left( 1+q \right) ^{2}}}
-{\frac {{{\chi_{2x}}}^{2}}{ \left( 1+q \right) ^{2}}}
\right. \cr && \left.
+\frac{1}{2}\,{\frac {{{\chi_{2y}}}^{2}}{ \left( 1+q \right) ^{2}}}
+\frac{1}{2}\,{\frac {{{\chi_{2z}}}^{2}}{ \left( 1+q \right) ^{2}}}
-{\frac {18\,{q}^{4}-9\,{q}^{3}-62\,{q}^{2}-9\,q+18}
{72\, \left( 1+q \right) ^{4}}} \right)  \left( M\Omega \right) ^{2/3}
\cr &&
+ \left( -\frac{1}{24}\,{\frac {q \left( 26\,{q}^{2}+6\,q-3 \right) {\chi_{1z}}}
{ \left( 1+q \right) ^{4}}}
+\frac{1}{24}\,{\frac {q \left( 3\,{q}^{2}-6\,q-26 \right) 
{\chi_{2z}}}{ \left( 1+q \right) ^{4}}} \right) M\Omega
\cr &&
+ \left( -\frac{1}{24}\,{\frac {{q}^{2} \left( 8\,{q}^{2}-40\,q-71 \right) 
{{\chi_{1x}}}^{2}}{ \left( 1+q \right) ^{4}}}
+\frac{1}{12}\,{\frac {q \left( 36\,{q}^{2}+47\,q+36 \right) {\chi_{2x}}\,{\chi_{1x}}}
{ \left( 1+q \right) ^{4}}}
-\frac{1}{6}\,{\frac {{q}^{2} \left( 11\,{q}^{2}+25\,q+17 \right) {{\chi_{1y}}}^{2}}
{ \left( 1+q \right) ^{4}}}
\right. \cr && \left.
-\frac{1}{2}\,{\frac {q \left( 11\,{q}^{2}+20\,q+11 \right) {\chi_{2y}}\,{\chi_{1y}}}
{ \left( 1+q \right) ^{4}}}
+\frac{1}{18}\,{\frac {{q}^{2} \left( 7\,{q}^{2}-15\,q-27 \right) {{\chi_{1z}}}^{2}}
{ \left( 1+q \right) ^{4}}}
-\frac{1}{9}\,{\frac {q \left( 15\,{q}^{2}+17\,q+15 \right) {\chi_{2z}}\,{\chi_{1z}}}
{ \left( 1+q \right) ^{4}}}
\right. \cr && \left.
+\frac{1}{24}\,{\frac { \left( 71\,{q}^{2}+40\,q-8 \right) {{\chi_{2x}}}^{2}}
{ \left( 1+q \right) ^{4}}}
-\frac{1}{6}\,{\frac { \left( 17\,{q}^{2}+25\,q+11 \right) {{\chi_{2y}}}^{2}}
{ \left( 1+q \right) ^{4}}}
-\frac{1}{18}\,{\frac { \left( 27\,{q}^{2}+15\,q-7 \right) {{\chi_{2z}}}^{2}}
{ \left( 1+q \right) ^{4}}}
+{\frac {167\,{\pi}^{2}q}{192\, \left( 1+q \right) ^{2}}}
\right. \cr && \left.
-{\frac {324\,{q}^{6}+16569\,{q}^{5}+65304\,{q}^{4}+98086\,{q}^{3}
+65304\,{q}^{2}+16569\,q+324}{1296\, \left( 1+q \right) ^{6}}} \right)
\left( M\Omega \right) ^{4/3} \,,
\label{eq:qc_radius}
\\
\frac{P_t}{M} &=&
\frac {q}{ \left( 1+q \right) ^{2}}
\biggl[ \left( M\Omega \right)^{1/3}
+\frac{1}{6}\,{\frac { \left( 15\,{q}^{2}+29\,q+15 \right) M\Omega}
{ \left( 1+q \right) ^{2}}} 
+ \left( -\frac{2}{3}\,{\frac { \left( 3+4\,q \right) q{\chi_{1z}}}
{ \left( 1+q \right) ^{2}}}
-\frac{2}{3}\,{\frac { \left( 3\,q+4 \right) {\chi_{2z}}}
{ \left( 1+q \right) ^{2}}} \right)  \left( M\Omega \right) ^{4/3}
\cr &&
+ \left( -{\frac {{{\chi_{1x}}}^{2}{q}^{2}}{ \left( 1+q \right) ^{2}}}
-2\,{\frac {{\chi_{1x}}\,{\chi_{2x}}\,q}{ \left( 1+q \right) ^{2}}}
+\frac{1}{2}\,{\frac {{{\chi_{1y}}}^{2}{q}^{2}}{ \left( 1+q \right) ^{2}}}
+{\frac {{\chi_{1y}}\,{\chi_{2y}}\,q}{ \left( 1+q \right) ^{2}}}
+\frac{1}{2}\,{\frac {{{\chi_{1z}}}^{2}{q}^{2}}{ \left( 1+q \right) ^{2}}}
+{\frac {{\chi_{1z}}\,{\chi_{2z}}\,q}{ \left( 1+q \right) ^{2}}}
-{\frac {{{\chi_{2x}}}^{2}}{ \left( 1+q \right) ^{2}}}
\right. \cr && \left.
+\frac{1}{2}\,{\frac {{{\chi_{2y}}}^{2}}{ \left( 1+q \right) ^{2}}}
+\frac{1}{2}\,{\frac {{{\chi_{2z}}}^{2}}{ \left( 1+q \right) ^{2}}}
+{\frac {441\,{q}^{4}+1440\,{q}^{3}+1997\,{q}^{2}+1440\,q+441}
{72\, \left( 1+q \right) ^{4}}} \right)  \left( M\Omega \right) ^{5/3}
\cr &&
+ \left( -\frac{1}{2}\,{\frac {q \left( 16\,{q}^{3}+29\,{q}^{2}+22\,q+7 \right) 
{\chi_{1z}}}{ \left( 1+q \right) ^{4}}}
-\frac{1}{2}\,{\frac { \left( 7\,{q}^{3}+22\,{q}^{2}+29\,q+16 \right) {\chi_{2z}}}
{ \left( 1+q \right) ^{4}}} \right)  \left( M\Omega \right)^{2}
\cr &&
+\left( -\frac{1}{24}\,{\frac {{q}^{2} \left( 116\,{q}^{2}-4\,q-53 \right) 
{{\chi_{1x}}}^{2}}{ \left( 1+q \right) ^{4}}}
+{\frac {53\,{\chi_{1x}}\,{\chi_{2x}}\,{q}^{2}}
{12\, \left( 1+q \right) ^{4}}}
+\frac{1}{12}\,{\frac {{q}^{2} \left( 5\,{q}^{2}-41\,q-31 \right) {{\chi_{1y}}}^{2}}
{ \left( 1+q \right) ^{4}}}
\right. \cr && \left.
-\frac{1}{2}\,{\frac {q \left( 8\,{q}^{2}+21\,q+8 \right) {\chi_{2y}}\,{\chi_{1y}}}
{ \left( 1+q \right) ^{4}}}
-\frac{1}{36}\,{\frac {{q}^{2} \left( {q}^{2}+147\,q+81 \right) {{\chi_{1z}}}^{2}}
{ \left( 1+q \right) ^{4}}}
-\frac{1}{18}\,{\frac {q \left( 21\,{q}^{2}+67\,q+21 \right) {\chi_{2z}}\,{\chi_{1z}}}
{ \left( 1+q \right) ^{4}}}
\right. \cr && \left.
+\frac{1}{24}\,{\frac { \left( 53\,{q}^{2}+4\,q-116 \right) {{\chi_{2x}}}^{2}}
{ \left( 1+q \right) ^{4}}}
-\frac{1}{12}\,{\frac { \left( 31\,{q}^{2}+41\,q-5 \right) {{\chi_{2y}}}^{2}}
{ \left( 1+q \right) ^{4}}}
-\frac{1}{36}\,{\frac { \left( 81\,{q}^{2}+147\,q+1 \right) {{\chi_{2z}}}^{2}}
{ \left( 1+q \right) ^{4}}}
+{\frac {161\,{\pi}^{2}q}{192\, \left( 1+q \right) ^{2}}}
\right. \cr && \left.
+{\frac {20007\,{q}^{6}+60489\,{q}^{5}+67320\,{q}^{4}+53681\,{q}^{3}
+67320\,{q}^{2}+60489\,q+20007}{1296\, \left( 1+q \right) ^{6}}} \right)  
\left( M\Omega \right) ^{7/3} 
\biggr] \,.
\eea
\end{widetext}


\bibliographystyle{apsrev4-1}
\bibliography{../../Bibtex/references}

\begin{thebibliography}{72}%
\makeatletter
\providecommand \@ifxundefined [1]{%
 \@ifx{#1\undefined}
}%
\providecommand \@ifnum [1]{%
 \ifnum #1\expandafter \@firstoftwo
 \else \expandafter \@secondoftwo
 \fi
}%
\providecommand \@ifx [1]{%
 \ifx #1\expandafter \@firstoftwo
 \else \expandafter \@secondoftwo
 \fi
}%
\providecommand \natexlab [1]{#1}%
\providecommand \enquote  [1]{``#1''}%
\providecommand \bibnamefont  [1]{#1}%
\providecommand \bibfnamefont [1]{#1}%
\providecommand \citenamefont [1]{#1}%
\providecommand \href@noop [0]{\@secondoftwo}%
\providecommand \href [0]{\begingroup \@sanitize@url \@href}%
\providecommand \@href[1]{\@@startlink{#1}\@@href}%
\providecommand \@@href[1]{\endgroup#1\@@endlink}%
\providecommand \@sanitize@url [0]{\catcode `\\12\catcode `\$12\catcode
  `\&12\catcode `\#12\catcode `\^12\catcode `\_12\catcode `\%12\relax}%
\providecommand \@@startlink[1]{}%
\providecommand \@@endlink[0]{}%
\providecommand \url  [0]{\begingroup\@sanitize@url \@url }%
\providecommand \@url [1]{\endgroup\@href {#1}{\urlprefix }}%
\providecommand \urlprefix  [0]{URL }%
\providecommand \Eprint [0]{\href }%
\providecommand \doibase [0]{http://dx.doi.org/}%
\providecommand \selectlanguage [0]{\@gobble}%
\providecommand \bibinfo  [0]{\@secondoftwo}%
\providecommand \bibfield  [0]{\@secondoftwo}%
\providecommand \translation [1]{[#1]}%
\providecommand \BibitemOpen [0]{}%
\providecommand \bibitemStop [0]{}%
\providecommand \bibitemNoStop [0]{.\EOS\space}%
\providecommand \EOS [0]{\spacefactor3000\relax}%
\providecommand \BibitemShut  [1]{\csname bibitem#1\endcsname}%
\let\auto@bib@innerbib\@empty
\bibitem [{\citenamefont {Pretorius}(2005)}]{Pretorius:2005gq}%
  \BibitemOpen
  \bibfield  {author} {\bibinfo {author} {\bibfnamefont {F.}~\bibnamefont
  {Pretorius}},\ }\href@noop {} {\bibfield  {journal} {\bibinfo  {journal}
  {Phys. Rev. Lett.}\ }\textbf {\bibinfo {volume} {95}},\ \bibinfo {pages}
  {121101} (\bibinfo {year} {2005})},\ \Eprint
  {http://arxiv.org/abs/gr-qc/0507014} {gr-qc/0507014} \BibitemShut {NoStop}%
\bibitem [{\citenamefont {Campanelli}\ \emph
  {et~al.}(2006{\natexlab{a}})\citenamefont {Campanelli}, \citenamefont
  {Lousto}, \citenamefont {Marronetti},\ and\ \citenamefont
  {Zlochower}}]{Campanelli:2005dd}%
  \BibitemOpen
  \bibfield  {author} {\bibinfo {author} {\bibfnamefont {M.}~\bibnamefont
  {Campanelli}}, \bibinfo {author} {\bibfnamefont {C.~O.}\ \bibnamefont
  {Lousto}}, \bibinfo {author} {\bibfnamefont {P.}~\bibnamefont {Marronetti}},
  \ and\ \bibinfo {author} {\bibfnamefont {Y.}~\bibnamefont {Zlochower}},\
  }\href@noop {} {\bibfield  {journal} {\bibinfo  {journal} {Phys. Rev. Lett.}\
  }\textbf {\bibinfo {volume} {96}},\ \bibinfo {pages} {111101} (\bibinfo
  {year} {2006}{\natexlab{a}})},\ \Eprint {http://arxiv.org/abs/gr-qc/0511048}
  {gr-qc/0511048} \BibitemShut {NoStop}%
\bibitem [{\citenamefont {Baker}\ \emph {et~al.}(2006)\citenamefont {Baker},
  \citenamefont {Centrella}, \citenamefont {Choi}, \citenamefont {Koppitz},\
  and\ \citenamefont {van Meter}}]{Baker:2005vv}%
  \BibitemOpen
  \bibfield  {author} {\bibinfo {author} {\bibfnamefont {J.~G.}\ \bibnamefont
  {Baker}}, \bibinfo {author} {\bibfnamefont {J.}~\bibnamefont {Centrella}},
  \bibinfo {author} {\bibfnamefont {D.-I.}\ \bibnamefont {Choi}}, \bibinfo
  {author} {\bibfnamefont {M.}~\bibnamefont {Koppitz}}, \ and\ \bibinfo
  {author} {\bibfnamefont {J.}~\bibnamefont {van Meter}},\ }\href@noop {}
  {\bibfield  {journal} {\bibinfo  {journal} {Phys. Rev. Lett.}\ }\textbf
  {\bibinfo {volume} {96}},\ \bibinfo {pages} {111102} (\bibinfo {year}
  {2006})},\ \Eprint {http://arxiv.org/abs/gr-qc/0511103} {gr-qc/0511103}
  \BibitemShut {NoStop}%
\bibitem [{\citenamefont {Abbott}\ \emph
  {et~al.}(2016{\natexlab{a}})\citenamefont {Abbott} \emph
  {et~al.}}]{Abbott:2016blz}%
  \BibitemOpen
  \bibfield  {author} {\bibinfo {author} {\bibfnamefont {B.}~\bibnamefont
  {Abbott}} \emph {et~al.} (\bibinfo {collaboration} {Virgo, LIGO
  Scientific}),\ }\href {\doibase 10.1103/PhysRevLett.116.061102} {\bibfield
  {journal} {\bibinfo  {journal} {Phys. Rev. Lett.}\ }\textbf {\bibinfo
  {volume} {116}},\ \bibinfo {pages} {061102} (\bibinfo {year}
  {2016}{\natexlab{a}})},\ \Eprint {http://arxiv.org/abs/1602.03837}
  {arXiv:1602.03837 [gr-qc]} \BibitemShut {NoStop}%
\bibitem [{\citenamefont {Abbott}\ \emph
  {et~al.}(2016{\natexlab{b}})\citenamefont {Abbott} \emph
  {et~al.}}]{Abbott:2016nmj}%
  \BibitemOpen
  \bibfield  {author} {\bibinfo {author} {\bibfnamefont {B.~P.}\ \bibnamefont
  {Abbott}} \emph {et~al.} (\bibinfo {collaboration} {Virgo, LIGO
  Scientific}),\ }\href {\doibase 10.1103/PhysRevLett.116.241103} {\bibfield
  {journal} {\bibinfo  {journal} {Phys. Rev. Lett.}\ }\textbf {\bibinfo
  {volume} {116}},\ \bibinfo {pages} {241103} (\bibinfo {year}
  {2016}{\natexlab{b}})},\ \Eprint {http://arxiv.org/abs/1606.04855}
  {arXiv:1606.04855 [gr-qc]} \BibitemShut {NoStop}%
\bibitem [{\citenamefont {Abbott}\ \emph
  {et~al.}(2016{\natexlab{c}})\citenamefont {Abbott} \emph
  {et~al.}}]{Abbott:2016apu}%
  \BibitemOpen
  \bibfield  {author} {\bibinfo {author} {\bibfnamefont {B.~P.}\ \bibnamefont
  {Abbott}} \emph {et~al.} (\bibinfo {collaboration} {Virgo, LIGO
  Scientific}),\ }\href {\doibase 10.1103/PhysRevD.94.064035} {\bibfield
  {journal} {\bibinfo  {journal} {Phys. Rev.}\ }\textbf {\bibinfo {volume}
  {D94}},\ \bibinfo {pages} {064035} (\bibinfo {year} {2016}{\natexlab{c}})},\
  \Eprint {http://arxiv.org/abs/1606.01262} {arXiv:1606.01262 [gr-qc]}
  \BibitemShut {NoStop}%
\bibitem [{\citenamefont {Abbott}\ \emph
  {et~al.}(2016{\natexlab{d}})\citenamefont {Abbott} \emph
  {et~al.}}]{TheLIGOScientific:2016wfe}%
  \BibitemOpen
  \bibfield  {author} {\bibinfo {author} {\bibfnamefont {B.~P.}\ \bibnamefont
  {Abbott}} \emph {et~al.} (\bibinfo {collaboration} {Virgo, LIGO
  Scientific}),\ }\href {\doibase 10.1103/PhysRevLett.116.241102} {\bibfield
  {journal} {\bibinfo  {journal} {Phys. Rev. Lett.}\ }\textbf {\bibinfo
  {volume} {116}},\ \bibinfo {pages} {241102} (\bibinfo {year}
  {2016}{\natexlab{d}})},\ \Eprint {http://arxiv.org/abs/1602.03840}
  {arXiv:1602.03840 [gr-qc]} \BibitemShut {NoStop}%
\bibitem [{\citenamefont {Abbott}\ \emph
  {et~al.}(2016{\natexlab{e}})\citenamefont {Abbott} \emph
  {et~al.}}]{TheLIGOScientific:2016src}%
  \BibitemOpen
  \bibfield  {author} {\bibinfo {author} {\bibfnamefont {B.~P.}\ \bibnamefont
  {Abbott}} \emph {et~al.} (\bibinfo {collaboration} {Virgo, LIGO
  Scientific}),\ }\href {\doibase 10.1103/PhysRevLett.116.221101} {\bibfield
  {journal} {\bibinfo  {journal} {Phys. Rev. Lett.}\ }\textbf {\bibinfo
  {volume} {116}},\ \bibinfo {pages} {221101} (\bibinfo {year}
  {2016}{\natexlab{e}})},\ \Eprint {http://arxiv.org/abs/1602.03841}
  {arXiv:1602.03841 [gr-qc]} \BibitemShut {NoStop}%
\bibitem [{\citenamefont {Abbott}\ \emph
  {et~al.}(2016{\natexlab{f}})\citenamefont {Abbott} \emph
  {et~al.}}]{Abbott:2016izl}%
  \BibitemOpen
  \bibfield  {author} {\bibinfo {author} {\bibfnamefont {B.~P.}\ \bibnamefont
  {Abbott}} \emph {et~al.} (\bibinfo {collaboration} {Virgo, LIGO
  Scientific}),\ }\href {\doibase 10.1103/PhysRevX.6.041014} {\bibfield
  {journal} {\bibinfo  {journal} {Phys. Rev.}\ }\textbf {\bibinfo {volume}
  {X6}},\ \bibinfo {pages} {041014} (\bibinfo {year} {2016}{\natexlab{f}})},\
  \Eprint {http://arxiv.org/abs/1606.01210} {arXiv:1606.01210 [gr-qc]}
  \BibitemShut {NoStop}%
\bibitem [{\citenamefont {Abbott}\ \emph
  {et~al.}(2016{\natexlab{g}})\citenamefont {Abbott} \emph
  {et~al.}}]{TheLIGOScientific:2016pea}%
  \BibitemOpen
  \bibfield  {author} {\bibinfo {author} {\bibfnamefont {B.~P.}\ \bibnamefont
  {Abbott}} \emph {et~al.} (\bibinfo {collaboration} {Virgo, LIGO
  Scientific}),\ }\href {\doibase 10.1103/PhysRevX.6.041015} {\bibfield
  {journal} {\bibinfo  {journal} {Phys. Rev.}\ }\textbf {\bibinfo {volume}
  {X6}},\ \bibinfo {pages} {041015} (\bibinfo {year} {2016}{\natexlab{g}})},\
  \Eprint {http://arxiv.org/abs/1606.04856} {arXiv:1606.04856 [gr-qc]}
  \BibitemShut {NoStop}%
\bibitem [{\citenamefont {Lovelace}\ \emph {et~al.}(2016)\citenamefont
  {Lovelace} \emph {et~al.}}]{Lovelace:2016uwp}%
  \BibitemOpen
  \bibfield  {author} {\bibinfo {author} {\bibfnamefont {G.}~\bibnamefont
  {Lovelace}} \emph {et~al.},\ }\href {\doibase 10.1088/0264-9381/33/24/244002}
  {\bibfield  {journal} {\bibinfo  {journal} {Class. Quant. Grav.}\ }\textbf
  {\bibinfo {volume} {33}},\ \bibinfo {pages} {244002} (\bibinfo {year}
  {2016})},\ \Eprint {http://arxiv.org/abs/1607.05377} {arXiv:1607.05377
  [gr-qc]} \BibitemShut {NoStop}%
\bibitem [{\citenamefont {Peters}(1964)}]{Peters:1964zz}%
  \BibitemOpen
  \bibfield  {author} {\bibinfo {author} {\bibfnamefont {P.}~\bibnamefont
  {Peters}},\ }\href {\doibase 10.1103/PhysRev.136.B1224} {\bibfield  {journal}
  {\bibinfo  {journal} {Phys. Rev.}\ }\textbf {\bibinfo {volume} {136}},\
  \bibinfo {pages} {B1224} (\bibinfo {year} {1964})}\BibitemShut {NoStop}%
\bibitem [{\citenamefont {Mroue}\ \emph {et~al.}(2010)\citenamefont {Mroue},
  \citenamefont {Pfeiffer}, \citenamefont {Kidder},\ and\ \citenamefont
  {Teukolsky}}]{Mroue:2010re}%
  \BibitemOpen
  \bibfield  {author} {\bibinfo {author} {\bibfnamefont {A.~H.}\ \bibnamefont
  {Mroue}}, \bibinfo {author} {\bibfnamefont {H.~P.}\ \bibnamefont {Pfeiffer}},
  \bibinfo {author} {\bibfnamefont {L.~E.}\ \bibnamefont {Kidder}}, \ and\
  \bibinfo {author} {\bibfnamefont {S.~A.}\ \bibnamefont {Teukolsky}},\ }\href
  {\doibase 10.1103/PhysRevD.82.124016} {\bibfield  {journal} {\bibinfo
  {journal} {Phys. Rev.}\ }\textbf {\bibinfo {volume} {D82}},\ \bibinfo {pages}
  {124016} (\bibinfo {year} {2010})},\ \Eprint {http://arxiv.org/abs/1004.4697}
  {arXiv:1004.4697 [gr-qc]} \BibitemShut {NoStop}%
\bibitem [{\citenamefont {Lousto}\ \emph {et~al.}(2016)\citenamefont {Lousto},
  \citenamefont {Healy},\ and\ \citenamefont {Nakano}}]{Lousto:2015uwa}%
  \BibitemOpen
  \bibfield  {author} {\bibinfo {author} {\bibfnamefont {C.~O.}\ \bibnamefont
  {Lousto}}, \bibinfo {author} {\bibfnamefont {J.}~\bibnamefont {Healy}}, \
  and\ \bibinfo {author} {\bibfnamefont {H.}~\bibnamefont {Nakano}},\ }\href
  {\doibase 10.1103/PhysRevD.93.044031} {\bibfield  {journal} {\bibinfo
  {journal} {Phys. Rev.}\ }\textbf {\bibinfo {volume} {D93}},\ \bibinfo {pages}
  {044031} (\bibinfo {year} {2016})},\ \Eprint
  {http://arxiv.org/abs/1506.04768} {arXiv:1506.04768 [gr-qc]} \BibitemShut
  {NoStop}%
\bibitem [{\citenamefont {Lousto}\ and\ \citenamefont
  {Zlochower}(2013)}]{Lousto:2013oza}%
  \BibitemOpen
  \bibfield  {author} {\bibinfo {author} {\bibfnamefont {C.~O.}\ \bibnamefont
  {Lousto}}\ and\ \bibinfo {author} {\bibfnamefont {Y.}~\bibnamefont
  {Zlochower}},\ }\href {\doibase 10.1103/PhysRevD.88.024001} {\bibfield
  {journal} {\bibinfo  {journal} {Phys. Rev.}\ }\textbf {\bibinfo {volume}
  {D88}},\ \bibinfo {pages} {024001} (\bibinfo {year} {2013})},\ \Eprint
  {http://arxiv.org/abs/1304.3937} {arXiv:1304.3937 [gr-qc]} \BibitemShut
  {NoStop}%
\bibitem [{\citenamefont {Lousto}\ and\ \citenamefont
  {Healy}(2015)}]{Lousto:2014ida}%
  \BibitemOpen
  \bibfield  {author} {\bibinfo {author} {\bibfnamefont {C.~O.}\ \bibnamefont
  {Lousto}}\ and\ \bibinfo {author} {\bibfnamefont {J.}~\bibnamefont {Healy}},\
  }\href {\doibase 10.1103/PhysRevLett.114.141101} {\bibfield  {journal}
  {\bibinfo  {journal} {Phys. Rev. Lett.}\ }\textbf {\bibinfo {volume} {114}},\
  \bibinfo {pages} {141101} (\bibinfo {year} {2015})},\ \Eprint
  {http://arxiv.org/abs/1410.3830} {arXiv:1410.3830 [gr-qc]} \BibitemShut
  {NoStop}%
\bibitem [{\citenamefont {Szilagyi}\ \emph {et~al.}(2015)\citenamefont
  {Szilagyi}, \citenamefont {Blackman}, \citenamefont {Buonanno}, \citenamefont
  {Taracchini}, \citenamefont {Pfeiffer}, \citenamefont {Scheel}, \citenamefont
  {Chu}, \citenamefont {Kidder},\ and\ \citenamefont {Pan}}]{Szilagyi:2015rwa}%
  \BibitemOpen
  \bibfield  {author} {\bibinfo {author} {\bibfnamefont {B.}~\bibnamefont
  {Szilagyi}}, \bibinfo {author} {\bibfnamefont {J.}~\bibnamefont {Blackman}},
  \bibinfo {author} {\bibfnamefont {A.}~\bibnamefont {Buonanno}}, \bibinfo
  {author} {\bibfnamefont {A.}~\bibnamefont {Taracchini}}, \bibinfo {author}
  {\bibfnamefont {H.~P.}\ \bibnamefont {Pfeiffer}}, \bibinfo {author}
  {\bibfnamefont {M.~A.}\ \bibnamefont {Scheel}}, \bibinfo {author}
  {\bibfnamefont {T.}~\bibnamefont {Chu}}, \bibinfo {author} {\bibfnamefont
  {L.~E.}\ \bibnamefont {Kidder}}, \ and\ \bibinfo {author} {\bibfnamefont
  {Y.}~\bibnamefont {Pan}},\ }\href {\doibase 10.1103/PhysRevLett.115.031102}
  {\bibfield  {journal} {\bibinfo  {journal} {Phys. Rev. Lett.}\ }\textbf
  {\bibinfo {volume} {115}},\ \bibinfo {pages} {031102} (\bibinfo {year}
  {2015})},\ \Eprint {http://arxiv.org/abs/1502.04953} {arXiv:1502.04953
  [gr-qc]} \BibitemShut {NoStop}%
\bibitem [{\citenamefont {Baker}\ \emph {et~al.}(2002)\citenamefont {Baker},
  \citenamefont {Campanelli}, \citenamefont {Lousto},\ and\ \citenamefont
  {Takahashi}}]{Baker:2002qf}%
  \BibitemOpen
  \bibfield  {author} {\bibinfo {author} {\bibfnamefont {J.~G.}\ \bibnamefont
  {Baker}}, \bibinfo {author} {\bibfnamefont {M.}~\bibnamefont {Campanelli}},
  \bibinfo {author} {\bibfnamefont {C.}~\bibnamefont {Lousto}}, \ and\ \bibinfo
  {author} {\bibfnamefont {R.}~\bibnamefont {Takahashi}},\ }\href {\doibase
  10.1103/PhysRevD.65.124012} {\bibfield  {journal} {\bibinfo  {journal} {Phys.
  Rev.}\ }\textbf {\bibinfo {volume} {D65}},\ \bibinfo {pages} {124012}
  (\bibinfo {year} {2002})},\ \Eprint {http://arxiv.org/abs/astro-ph/0202469}
  {arXiv:astro-ph/0202469 [astro-ph]} \BibitemShut {NoStop}%
\bibitem [{\citenamefont {Husa}\ \emph {et~al.}(2008)\citenamefont {Husa},
  \citenamefont {Hannam}, \citenamefont {Gonzalez}, \citenamefont {Sperhake},\
  and\ \citenamefont {Brugmann}}]{Husa:2007rh}%
  \BibitemOpen
  \bibfield  {author} {\bibinfo {author} {\bibfnamefont {S.}~\bibnamefont
  {Husa}}, \bibinfo {author} {\bibfnamefont {M.}~\bibnamefont {Hannam}},
  \bibinfo {author} {\bibfnamefont {J.~A.}\ \bibnamefont {Gonzalez}}, \bibinfo
  {author} {\bibfnamefont {U.}~\bibnamefont {Sperhake}}, \ and\ \bibinfo
  {author} {\bibfnamefont {B.}~\bibnamefont {Brugmann}},\ }\href {\doibase
  10.1103/PhysRevD.77.044037} {\bibfield  {journal} {\bibinfo  {journal} {Phys.
  Rev.}\ }\textbf {\bibinfo {volume} {D77}},\ \bibinfo {pages} {044037}
  (\bibinfo {year} {2008})},\ \Eprint {http://arxiv.org/abs/0706.0904}
  {arXiv:0706.0904 [gr-qc]} \BibitemShut {NoStop}%
\bibitem [{\citenamefont {Campanelli}\ \emph {et~al.}(2009)\citenamefont
  {Campanelli}, \citenamefont {Lousto}, \citenamefont {Nakano},\ and\
  \citenamefont {Zlochower}}]{Campanelli:2008nk}%
  \BibitemOpen
  \bibfield  {author} {\bibinfo {author} {\bibfnamefont {M.}~\bibnamefont
  {Campanelli}}, \bibinfo {author} {\bibfnamefont {C.~O.}\ \bibnamefont
  {Lousto}}, \bibinfo {author} {\bibfnamefont {H.}~\bibnamefont {Nakano}}, \
  and\ \bibinfo {author} {\bibfnamefont {Y.}~\bibnamefont {Zlochower}},\ }\href
  {\doibase 10.1103/PhysRevD.79.084010} {\bibfield  {journal} {\bibinfo
  {journal} {Phys. Rev.}\ }\textbf {\bibinfo {volume} {D79}},\ \bibinfo {pages}
  {084010} (\bibinfo {year} {2009})},\ \Eprint {http://arxiv.org/abs/0808.0713}
  {arXiv:0808.0713 [gr-qc]} \BibitemShut {NoStop}%
\bibitem [{\citenamefont {Pfeiffer}\ \emph {et~al.}(2007)\citenamefont
  {Pfeiffer}, \citenamefont {Brown}, \citenamefont {Kidder}, \citenamefont
  {Lindblom}, \citenamefont {Lovelace},\ and\ \citenamefont
  {Scheel}}]{Pfeiffer:2007yz}%
  \BibitemOpen
  \bibfield  {author} {\bibinfo {author} {\bibfnamefont {H.~P.}\ \bibnamefont
  {Pfeiffer}}, \bibinfo {author} {\bibfnamefont {D.~A.}\ \bibnamefont {Brown}},
  \bibinfo {author} {\bibfnamefont {L.~E.}\ \bibnamefont {Kidder}}, \bibinfo
  {author} {\bibfnamefont {L.}~\bibnamefont {Lindblom}}, \bibinfo {author}
  {\bibfnamefont {G.}~\bibnamefont {Lovelace}}, \ and\ \bibinfo {author}
  {\bibfnamefont {M.~A.}\ \bibnamefont {Scheel}},\ }\href {\doibase
  10.1088/0264-9381/24/12/S06} {\bibfield  {journal} {\bibinfo  {journal}
  {Class. Quant. Grav.}\ }\textbf {\bibinfo {volume} {24}},\ \bibinfo {pages}
  {S59} (\bibinfo {year} {2007})},\ \Eprint
  {http://arxiv.org/abs/gr-qc/0702106} {arXiv:gr-qc/0702106 [gr-qc]}
  \BibitemShut {NoStop}%
\bibitem [{\citenamefont {Buonanno}\ \emph {et~al.}(2011)\citenamefont
  {Buonanno}, \citenamefont {Kidder}, \citenamefont {Mroue}, \citenamefont
  {Pfeiffer},\ and\ \citenamefont {Taracchini}}]{Buonanno:2010yk}%
  \BibitemOpen
  \bibfield  {author} {\bibinfo {author} {\bibfnamefont {A.}~\bibnamefont
  {Buonanno}}, \bibinfo {author} {\bibfnamefont {L.~E.}\ \bibnamefont
  {Kidder}}, \bibinfo {author} {\bibfnamefont {A.~H.}\ \bibnamefont {Mroue}},
  \bibinfo {author} {\bibfnamefont {H.~P.}\ \bibnamefont {Pfeiffer}}, \ and\
  \bibinfo {author} {\bibfnamefont {A.}~\bibnamefont {Taracchini}},\ }\href
  {\doibase 10.1103/PhysRevD.83.104034} {\bibfield  {journal} {\bibinfo
  {journal} {Phys. Rev.}\ }\textbf {\bibinfo {volume} {D83}},\ \bibinfo {pages}
  {104034} (\bibinfo {year} {2011})},\ \Eprint {http://arxiv.org/abs/1012.1549}
  {arXiv:1012.1549 [gr-qc]} \BibitemShut {NoStop}%
\bibitem [{\citenamefont {Purrer}\ \emph {et~al.}(2012)\citenamefont {Purrer},
  \citenamefont {Husa},\ and\ \citenamefont {Hannam}}]{Purrer:2012wy}%
  \BibitemOpen
  \bibfield  {author} {\bibinfo {author} {\bibfnamefont {M.}~\bibnamefont
  {Purrer}}, \bibinfo {author} {\bibfnamefont {S.}~\bibnamefont {Husa}}, \ and\
  \bibinfo {author} {\bibfnamefont {M.}~\bibnamefont {Hannam}},\ }\href
  {\doibase 10.1103/PhysRevD.85.124051} {\bibfield  {journal} {\bibinfo
  {journal} {Phys. Rev.}\ }\textbf {\bibinfo {volume} {D85}},\ \bibinfo {pages}
  {124051} (\bibinfo {year} {2012})},\ \Eprint {http://arxiv.org/abs/1203.4258}
  {arXiv:1203.4258 [gr-qc]} \BibitemShut {NoStop}%
\bibitem [{\citenamefont {Buchman}\ \emph {et~al.}(2012)\citenamefont
  {Buchman}, \citenamefont {Pfeiffer}, \citenamefont {Scheel},\ and\
  \citenamefont {Szilagyi}}]{Buchman:2012dw}%
  \BibitemOpen
  \bibfield  {author} {\bibinfo {author} {\bibfnamefont {L.~T.}\ \bibnamefont
  {Buchman}}, \bibinfo {author} {\bibfnamefont {H.~P.}\ \bibnamefont
  {Pfeiffer}}, \bibinfo {author} {\bibfnamefont {M.~A.}\ \bibnamefont
  {Scheel}}, \ and\ \bibinfo {author} {\bibfnamefont {B.}~\bibnamefont
  {Szilagyi}},\ }\href {\doibase 10.1103/PhysRevD.86.084033} {\bibfield
  {journal} {\bibinfo  {journal} {Phys. Rev.}\ }\textbf {\bibinfo {volume}
  {D86}},\ \bibinfo {pages} {084033} (\bibinfo {year} {2012})},\ \Eprint
  {http://arxiv.org/abs/1206.3015} {arXiv:1206.3015 [gr-qc]} \BibitemShut
  {NoStop}%
\bibitem [{\citenamefont {Kelly}\ \emph {et~al.}(2007)\citenamefont {Kelly},
  \citenamefont {Tichy}, \citenamefont {Campanelli},\ and\ \citenamefont
  {Whiting}}]{Kelly:2007uc}%
  \BibitemOpen
  \bibfield  {author} {\bibinfo {author} {\bibfnamefont {B.~J.}\ \bibnamefont
  {Kelly}}, \bibinfo {author} {\bibfnamefont {W.}~\bibnamefont {Tichy}},
  \bibinfo {author} {\bibfnamefont {M.}~\bibnamefont {Campanelli}}, \ and\
  \bibinfo {author} {\bibfnamefont {B.~F.}\ \bibnamefont {Whiting}},\ }\href
  {\doibase 10.1103/PhysRevD.76.024008} {\bibfield  {journal} {\bibinfo
  {journal} {Phys. Rev.}\ }\textbf {\bibinfo {volume} {D76}},\ \bibinfo {pages}
  {024008} (\bibinfo {year} {2007})},\ \Eprint {http://arxiv.org/abs/0704.0628}
  {arXiv:0704.0628 [gr-qc]} \BibitemShut {NoStop}%
\bibitem [{\citenamefont {Tichy}\ \emph {et~al.}(2003)\citenamefont {Tichy},
  \citenamefont {Br{\"u}gmann}, \citenamefont {Campanelli},\ and\ \citenamefont
  {Diener}}]{Tichy:2002ec}%
  \BibitemOpen
  \bibfield  {author} {\bibinfo {author} {\bibfnamefont {W.}~\bibnamefont
  {Tichy}}, \bibinfo {author} {\bibfnamefont {B.}~\bibnamefont {Br{\"u}gmann}},
  \bibinfo {author} {\bibfnamefont {M.}~\bibnamefont {Campanelli}}, \ and\
  \bibinfo {author} {\bibfnamefont {P.}~\bibnamefont {Diener}},\ }\href@noop {}
  {\bibfield  {journal} {\bibinfo  {journal} {Phys. Rev.}\ }\textbf {\bibinfo
  {volume} {D67}},\ \bibinfo {pages} {064008} (\bibinfo {year} {2003})},\
  \Eprint {http://arxiv.org/abs/gr-qc/0207011} {gr-qc/0207011} \BibitemShut
  {NoStop}%
\bibitem [{\citenamefont {Brandt}\ and\ \citenamefont
  {Br{\"u}gmann}(1997)}]{Brandt97b}%
  \BibitemOpen
  \bibfield  {author} {\bibinfo {author} {\bibfnamefont {S.}~\bibnamefont
  {Brandt}}\ and\ \bibinfo {author} {\bibfnamefont {B.}~\bibnamefont
  {Br{\"u}gmann}},\ }\href@noop {} {\bibfield  {journal} {\bibinfo  {journal}
  {Phys. Rev. Lett.}\ }\textbf {\bibinfo {volume} {78}},\ \bibinfo {pages}
  {3606} (\bibinfo {year} {1997})},\ \Eprint
  {http://arxiv.org/abs/gr-qc/9703066} {gr-qc/9703066} \BibitemShut {NoStop}%
\bibitem [{\citenamefont {Buonanno}\ \emph {et~al.}(2006)\citenamefont
  {Buonanno}, \citenamefont {Chen},\ and\ \citenamefont
  {Damour}}]{Buonanno:2005xu}%
  \BibitemOpen
  \bibfield  {author} {\bibinfo {author} {\bibfnamefont {A.}~\bibnamefont
  {Buonanno}}, \bibinfo {author} {\bibfnamefont {Y.}~\bibnamefont {Chen}}, \
  and\ \bibinfo {author} {\bibfnamefont {T.}~\bibnamefont {Damour}},\
  }\href@noop {} {\bibfield  {journal} {\bibinfo  {journal} {Phys. Rev.}\
  }\textbf {\bibinfo {volume} {D74}},\ \bibinfo {pages} {104005} (\bibinfo
  {year} {2006})},\ \Eprint {http://arxiv.org/abs/gr-qc/0508067}
  {gr-qc/0508067} \BibitemShut {NoStop}%
\bibitem [{\citenamefont {Damour}\ \emph {et~al.}(2008)\citenamefont {Damour},
  \citenamefont {Jaranowski},\ and\ \citenamefont {Schafer}}]{Damour:2007nc}%
  \BibitemOpen
  \bibfield  {author} {\bibinfo {author} {\bibfnamefont {T.}~\bibnamefont
  {Damour}}, \bibinfo {author} {\bibfnamefont {P.}~\bibnamefont {Jaranowski}},
  \ and\ \bibinfo {author} {\bibfnamefont {G.}~\bibnamefont {Schafer}},\ }\href
  {\doibase 10.1103/PhysRevD.77.064032} {\bibfield  {journal} {\bibinfo
  {journal} {Phys. Rev.}\ }\textbf {\bibinfo {volume} {D77}},\ \bibinfo {pages}
  {064032} (\bibinfo {year} {2008})},\ \Eprint {http://arxiv.org/abs/0711.1048}
  {arXiv:0711.1048 [gr-qc]} \BibitemShut {NoStop}%
\bibitem [{\citenamefont {Steinhoff}\ \emph
  {et~al.}(2008{\natexlab{a}})\citenamefont {Steinhoff}, \citenamefont
  {Hergt},\ and\ \citenamefont {Schafer}}]{Steinhoff:2007mb}%
  \BibitemOpen
  \bibfield  {author} {\bibinfo {author} {\bibfnamefont {J.}~\bibnamefont
  {Steinhoff}}, \bibinfo {author} {\bibfnamefont {S.}~\bibnamefont {Hergt}}, \
  and\ \bibinfo {author} {\bibfnamefont {G.}~\bibnamefont {Schafer}},\
  }\href@noop {} {\bibfield  {journal} {\bibinfo  {journal} {Phys. Rev.}\
  }\textbf {\bibinfo {volume} {D77}},\ \bibinfo {pages} {081501(R)} (\bibinfo
  {year} {2008}{\natexlab{a}})},\ \Eprint {http://arxiv.org/abs/0712.1716}
  {arXiv:0712.1716 [gr-qc]} \BibitemShut {NoStop}%
\bibitem [{\citenamefont {Steinhoff}\ \emph
  {et~al.}(2008{\natexlab{b}})\citenamefont {Steinhoff}, \citenamefont
  {Schafer},\ and\ \citenamefont {Hergt}}]{Steinhoff:2008zr}%
  \BibitemOpen
  \bibfield  {author} {\bibinfo {author} {\bibfnamefont {J.}~\bibnamefont
  {Steinhoff}}, \bibinfo {author} {\bibfnamefont {G.}~\bibnamefont {Schafer}},
  \ and\ \bibinfo {author} {\bibfnamefont {S.}~\bibnamefont {Hergt}},\ }\href
  {\doibase 10.1103/PhysRevD.77.104018} {\bibfield  {journal} {\bibinfo
  {journal} {Phys. Rev.}\ }\textbf {\bibinfo {volume} {D77}},\ \bibinfo {pages}
  {104018} (\bibinfo {year} {2008}{\natexlab{b}})},\ \Eprint
  {http://arxiv.org/abs/0805.3136} {arXiv:0805.3136 [gr-qc]} \BibitemShut
  {NoStop}%
\bibitem [{\citenamefont {Steinhoff}\ \emph
  {et~al.}(2008{\natexlab{c}})\citenamefont {Steinhoff}, \citenamefont
  {Hergt},\ and\ \citenamefont {Schafer}}]{Steinhoff:2008ji}%
  \BibitemOpen
  \bibfield  {author} {\bibinfo {author} {\bibfnamefont {J.}~\bibnamefont
  {Steinhoff}}, \bibinfo {author} {\bibfnamefont {S.}~\bibnamefont {Hergt}}, \
  and\ \bibinfo {author} {\bibfnamefont {G.}~\bibnamefont {Schafer}},\ }\href
  {\doibase 10.1103/PhysRevD.78.101503} {\bibfield  {journal} {\bibinfo
  {journal} {Phys. Rev.}\ }\textbf {\bibinfo {volume} {D78}},\ \bibinfo {pages}
  {101503} (\bibinfo {year} {2008}{\natexlab{c}})},\ \Eprint
  {http://arxiv.org/abs/0809.2200} {arXiv:0809.2200 [gr-qc]} \BibitemShut
  {NoStop}%
\bibitem [{\citenamefont {Damour}\ \emph {et~al.}(2000)\citenamefont {Damour},
  \citenamefont {Jaranowski},\ and\ \citenamefont {Schaefer}}]{Damour:2000kk}%
  \BibitemOpen
  \bibfield  {author} {\bibinfo {author} {\bibfnamefont {T.}~\bibnamefont
  {Damour}}, \bibinfo {author} {\bibfnamefont {P.}~\bibnamefont {Jaranowski}},
  \ and\ \bibinfo {author} {\bibfnamefont {G.}~\bibnamefont {Schaefer}},\
  }\href {\doibase 10.1103/PhysRevD.63.029903, 10.1103/PhysRevD.62.021501}
  {\bibfield  {journal} {\bibinfo  {journal} {Phys. Rev.}\ }\textbf {\bibinfo
  {volume} {D62}},\ \bibinfo {pages} {021501} (\bibinfo {year} {2000})},\
  \bibinfo {note} {[Erratum: Phys. Rev.D63,029903(2001)]},\ \Eprint
  {http://arxiv.org/abs/gr-qc/0003051} {arXiv:gr-qc/0003051 [gr-qc]}
  \BibitemShut {NoStop}%
\bibitem [{\citenamefont {Hergt}\ and\ \citenamefont
  {Schaefer}(2008)}]{Hergt:2008jn}%
  \BibitemOpen
  \bibfield  {author} {\bibinfo {author} {\bibfnamefont {S.}~\bibnamefont
  {Hergt}}\ and\ \bibinfo {author} {\bibfnamefont {G.}~\bibnamefont
  {Schaefer}},\ }\href {\doibase 10.1103/PhysRevD.78.124004} {\bibfield
  {journal} {\bibinfo  {journal} {Phys. Rev.}\ }\textbf {\bibinfo {volume}
  {D78}},\ \bibinfo {pages} {124004} (\bibinfo {year} {2008})},\ \Eprint
  {http://arxiv.org/abs/0809.2208} {arXiv:0809.2208 [gr-qc]} \BibitemShut
  {NoStop}%
\bibitem [{\citenamefont {{Ajith}}\ \emph {et~al.}(2007)\citenamefont
  {{Ajith}}, \citenamefont {{Boyle}}, \citenamefont {{Brown}}, \citenamefont
  {{Fairhurst}}, \citenamefont {{Hannam}}, \citenamefont {{Hinder}},
  \citenamefont {{Husa}}, \citenamefont {{Krishnan}}, \citenamefont {{Mercer}},
  \citenamefont {{Ohme}}, \citenamefont {{Ott}}, \citenamefont {{Read}},
  \citenamefont {{Santamaria}},\ and\ \citenamefont {{Whelan}}}]{Brown:2007jx}%
  \BibitemOpen
  \bibfield  {author} {\bibinfo {author} {\bibfnamefont {P.}~\bibnamefont
  {{Ajith}}}, \bibinfo {author} {\bibfnamefont {M.}~\bibnamefont {{Boyle}}},
  \bibinfo {author} {\bibfnamefont {D.~A.}\ \bibnamefont {{Brown}}}, \bibinfo
  {author} {\bibfnamefont {S.}~\bibnamefont {{Fairhurst}}}, \bibinfo {author}
  {\bibfnamefont {M.}~\bibnamefont {{Hannam}}}, \bibinfo {author}
  {\bibfnamefont {I.}~\bibnamefont {{Hinder}}}, \bibinfo {author}
  {\bibfnamefont {S.}~\bibnamefont {{Husa}}}, \bibinfo {author} {\bibfnamefont
  {B.}~\bibnamefont {{Krishnan}}}, \bibinfo {author} {\bibfnamefont {R.~A.}\
  \bibnamefont {{Mercer}}}, \bibinfo {author} {\bibfnamefont {F.}~\bibnamefont
  {{Ohme}}}, \bibinfo {author} {\bibfnamefont {C.~D.}\ \bibnamefont {{Ott}}},
  \bibinfo {author} {\bibfnamefont {J.~S.}\ \bibnamefont {{Read}}}, \bibinfo
  {author} {\bibfnamefont {L.}~\bibnamefont {{Santamaria}}}, \ and\ \bibinfo
  {author} {\bibfnamefont {J.~T.}\ \bibnamefont {{Whelan}}},\ }\href@noop {} {\
   (\bibinfo {year} {2007})},\ \Eprint {http://arxiv.org/abs/0709.0093}
  {arXiv:0709.0093 [gr-qc]} \BibitemShut {NoStop}%
\bibitem [{\citenamefont {Ossokine}\ \emph {et~al.}(2015)\citenamefont
  {Ossokine}, \citenamefont {Boyle}, \citenamefont {Kidder}, \citenamefont
  {Pfeiffer}, \citenamefont {Scheel},\ and\ \citenamefont
  {Szilagyi}}]{Ossokine:2015vda}%
  \BibitemOpen
  \bibfield  {author} {\bibinfo {author} {\bibfnamefont {S.}~\bibnamefont
  {Ossokine}}, \bibinfo {author} {\bibfnamefont {M.}~\bibnamefont {Boyle}},
  \bibinfo {author} {\bibfnamefont {L.~E.}\ \bibnamefont {Kidder}}, \bibinfo
  {author} {\bibfnamefont {H.~P.}\ \bibnamefont {Pfeiffer}}, \bibinfo {author}
  {\bibfnamefont {M.~A.}\ \bibnamefont {Scheel}}, \ and\ \bibinfo {author}
  {\bibfnamefont {B.}~\bibnamefont {Szilagyi}},\ }\href {\doibase
  10.1103/PhysRevD.92.104028} {\bibfield  {journal} {\bibinfo  {journal} {Phys.
  Rev.}\ }\textbf {\bibinfo {volume} {D92}},\ \bibinfo {pages} {104028}
  (\bibinfo {year} {2015})},\ \Eprint {http://arxiv.org/abs/1502.01747}
  {arXiv:1502.01747 [gr-qc]} \BibitemShut {NoStop}%
\bibitem [{\citenamefont {Alvi}(2001)}]{Alvi:2001mx}%
  \BibitemOpen
  \bibfield  {author} {\bibinfo {author} {\bibfnamefont {K.}~\bibnamefont
  {Alvi}},\ }\href {\doibase 10.1103/PhysRevD.64.104020} {\bibfield  {journal}
  {\bibinfo  {journal} {Phys. Rev.}\ }\textbf {\bibinfo {volume} {D64}},\
  \bibinfo {pages} {104020} (\bibinfo {year} {2001})},\ \Eprint
  {http://arxiv.org/abs/gr-qc/0107080} {arXiv:gr-qc/0107080 [gr-qc]}
  \BibitemShut {NoStop}%
\bibitem [{\citenamefont {Chatziioannou}\ \emph {et~al.}(2013)\citenamefont
  {Chatziioannou}, \citenamefont {Poisson},\ and\ \citenamefont
  {Yunes}}]{Chatziioannou:2012gq}%
  \BibitemOpen
  \bibfield  {author} {\bibinfo {author} {\bibfnamefont {K.}~\bibnamefont
  {Chatziioannou}}, \bibinfo {author} {\bibfnamefont {E.}~\bibnamefont
  {Poisson}}, \ and\ \bibinfo {author} {\bibfnamefont {N.}~\bibnamefont
  {Yunes}},\ }\href {\doibase 10.1103/PhysRevD.87.044022} {\bibfield  {journal}
  {\bibinfo  {journal} {Phys. Rev.}\ }\textbf {\bibinfo {volume} {D87}},\
  \bibinfo {pages} {044022} (\bibinfo {year} {2013})},\ \Eprint
  {http://arxiv.org/abs/1211.1686} {arXiv:1211.1686 [gr-qc]} \BibitemShut
  {NoStop}%
\bibitem [{\citenamefont {Arun}\ \emph {et~al.}(2009)\citenamefont {Arun},
  \citenamefont {Buonanno}, \citenamefont {Faye},\ and\ \citenamefont
  {Ochsner}}]{Arun:2008kb}%
  \BibitemOpen
  \bibfield  {author} {\bibinfo {author} {\bibfnamefont {K.~G.}\ \bibnamefont
  {Arun}}, \bibinfo {author} {\bibfnamefont {A.}~\bibnamefont {Buonanno}},
  \bibinfo {author} {\bibfnamefont {G.}~\bibnamefont {Faye}}, \ and\ \bibinfo
  {author} {\bibfnamefont {E.}~\bibnamefont {Ochsner}},\ }\href {\doibase
  10.1103/PhysRevD.79.104023} {\bibfield  {journal} {\bibinfo  {journal} {Phys.
  Rev.}\ }\textbf {\bibinfo {volume} {D79}},\ \bibinfo {pages} {104023}
  (\bibinfo {year} {2009})},\ \Eprint {http://arxiv.org/abs/0810.5336}
  {arXiv:0810.5336 [gr-qc]} \BibitemShut {NoStop}%
\bibitem [{\citenamefont {Blanchet}\ \emph {et~al.}(2011)\citenamefont
  {Blanchet}, \citenamefont {Buonanno},\ and\ \citenamefont
  {Faye}}]{Blanchet:2011zv}%
  \BibitemOpen
  \bibfield  {author} {\bibinfo {author} {\bibfnamefont {L.}~\bibnamefont
  {Blanchet}}, \bibinfo {author} {\bibfnamefont {A.}~\bibnamefont {Buonanno}},
  \ and\ \bibinfo {author} {\bibfnamefont {G.}~\bibnamefont {Faye}},\ }\href
  {\doibase 10.1103/PhysRevD.84.064041} {\bibfield  {journal} {\bibinfo
  {journal} {Phys. Rev.}\ }\textbf {\bibinfo {volume} {D84}},\ \bibinfo {pages}
  {064041} (\bibinfo {year} {2011})},\ \Eprint {http://arxiv.org/abs/1104.5659}
  {arXiv:1104.5659 [gr-qc]} \BibitemShut {NoStop}%
\bibitem [{\citenamefont {Pan}\ \emph {et~al.}(2010)\citenamefont {Pan} \emph
  {et~al.}}]{Pan:2009wj}%
  \BibitemOpen
  \bibfield  {author} {\bibinfo {author} {\bibfnamefont {Y.}~\bibnamefont
  {Pan}} \emph {et~al.},\ }\href {\doibase 10.1103/PhysRevD.81.084041}
  {\bibfield  {journal} {\bibinfo  {journal} {Phys. Rev.}\ }\textbf {\bibinfo
  {volume} {D81}},\ \bibinfo {pages} {084041} (\bibinfo {year} {2010})},\
  \Eprint {http://arxiv.org/abs/0912.3466} {arXiv:0912.3466 [gr-qc]}
  \BibitemShut {NoStop}%
\bibitem [{\citenamefont {Levi}\ and\ \citenamefont
  {Steinhoff}(2015)}]{Levi:2014gsa}%
  \BibitemOpen
  \bibfield  {author} {\bibinfo {author} {\bibfnamefont {M.}~\bibnamefont
  {Levi}}\ and\ \bibinfo {author} {\bibfnamefont {J.}~\bibnamefont
  {Steinhoff}},\ }\href {\doibase 10.1007/JHEP06(2015)059} {\bibfield
  {journal} {\bibinfo  {journal} {JHEP}\ }\textbf {\bibinfo {volume} {06}},\
  \bibinfo {pages} {059} (\bibinfo {year} {2015})},\ \Eprint
  {http://arxiv.org/abs/1410.2601} {arXiv:1410.2601 [gr-qc]} \BibitemShut
  {NoStop}%
\bibitem [{\citenamefont {Damour}\ \emph {et~al.}(2014)\citenamefont {Damour},
  \citenamefont {Jaranowski},\ and\ \citenamefont {Schafer}}]{Damour:2014jta}%
  \BibitemOpen
  \bibfield  {author} {\bibinfo {author} {\bibfnamefont {T.}~\bibnamefont
  {Damour}}, \bibinfo {author} {\bibfnamefont {P.}~\bibnamefont {Jaranowski}},
  \ and\ \bibinfo {author} {\bibfnamefont {G.}~\bibnamefont {Schafer}},\ }\href
  {\doibase 10.1103/PhysRevD.89.064058} {\bibfield  {journal} {\bibinfo
  {journal} {Phys. Rev.}\ }\textbf {\bibinfo {volume} {D89}},\ \bibinfo {pages}
  {064058} (\bibinfo {year} {2014})},\ \Eprint {http://arxiv.org/abs/1401.4548}
  {arXiv:1401.4548 [gr-qc]} \BibitemShut {NoStop}%
\bibitem [{\citenamefont {Damour}\ \emph {et~al.}(2015)\citenamefont {Damour},
  \citenamefont {Jaranowski},\ and\ \citenamefont
  {Sch\"afer}}]{Damour:2015isa}%
  \BibitemOpen
  \bibfield  {author} {\bibinfo {author} {\bibfnamefont {T.}~\bibnamefont
  {Damour}}, \bibinfo {author} {\bibfnamefont {P.}~\bibnamefont {Jaranowski}},
  \ and\ \bibinfo {author} {\bibfnamefont {G.}~\bibnamefont {Sch\"afer}},\
  }\href {\doibase 10.1103/PhysRevD.91.084024} {\bibfield  {journal} {\bibinfo
  {journal} {Phys. Rev.}\ }\textbf {\bibinfo {volume} {D91}},\ \bibinfo {pages}
  {084024} (\bibinfo {year} {2015})},\ \Eprint
  {http://arxiv.org/abs/1502.07245} {arXiv:1502.07245 [gr-qc]} \BibitemShut
  {NoStop}%
\bibitem [{\citenamefont {Bernard}\ \emph {et~al.}(2016)\citenamefont
  {Bernard}, \citenamefont {Blanchet}, \citenamefont {Bohe}, \citenamefont
  {Faye},\ and\ \citenamefont {Marsat}}]{Bernard:2016wrg}%
  \BibitemOpen
  \bibfield  {author} {\bibinfo {author} {\bibfnamefont {L.}~\bibnamefont
  {Bernard}}, \bibinfo {author} {\bibfnamefont {L.}~\bibnamefont {Blanchet}},
  \bibinfo {author} {\bibfnamefont {A.}~\bibnamefont {Bohe}}, \bibinfo {author}
  {\bibfnamefont {G.}~\bibnamefont {Faye}}, \ and\ \bibinfo {author}
  {\bibfnamefont {S.}~\bibnamefont {Marsat}},\ }\href@noop {} {\  (\bibinfo
  {year} {2016})},\ \Eprint {http://arxiv.org/abs/1610.07934} {arXiv:1610.07934
  [gr-qc]} \BibitemShut {NoStop}%
\bibitem [{\citenamefont {Hartung}\ \emph {et~al.}(2013)\citenamefont
  {Hartung}, \citenamefont {Steinhoff},\ and\ \citenamefont
  {Schafer}}]{Hartung:2013dza}%
  \BibitemOpen
  \bibfield  {author} {\bibinfo {author} {\bibfnamefont {J.}~\bibnamefont
  {Hartung}}, \bibinfo {author} {\bibfnamefont {J.}~\bibnamefont {Steinhoff}},
  \ and\ \bibinfo {author} {\bibfnamefont {G.}~\bibnamefont {Schafer}},\ }\href
  {\doibase 10.1002/andp.201200271} {\bibfield  {journal} {\bibinfo  {journal}
  {Annalen Phys.}\ }\textbf {\bibinfo {volume} {525}},\ \bibinfo {pages} {359}
  (\bibinfo {year} {2013})},\ \Eprint {http://arxiv.org/abs/1302.6723}
  {arXiv:1302.6723 [gr-qc]} \BibitemShut {NoStop}%
\bibitem [{\citenamefont {Levi}\ and\ \citenamefont
  {Steinhoff}(2016{\natexlab{a}})}]{Levi:2015uxa}%
  \BibitemOpen
  \bibfield  {author} {\bibinfo {author} {\bibfnamefont {M.}~\bibnamefont
  {Levi}}\ and\ \bibinfo {author} {\bibfnamefont {J.}~\bibnamefont
  {Steinhoff}},\ }\href {\doibase 10.1088/1475-7516/2016/01/011} {\bibfield
  {journal} {\bibinfo  {journal} {JCAP}\ }\textbf {\bibinfo {volume} {1601}},\
  \bibinfo {pages} {011} (\bibinfo {year} {2016}{\natexlab{a}})},\ \Eprint
  {http://arxiv.org/abs/1506.05056} {arXiv:1506.05056 [gr-qc]} \BibitemShut
  {NoStop}%
\bibitem [{\citenamefont {Levi}\ and\ \citenamefont
  {Steinhoff}(2014)}]{Levi:2014sba}%
  \BibitemOpen
  \bibfield  {author} {\bibinfo {author} {\bibfnamefont {M.}~\bibnamefont
  {Levi}}\ and\ \bibinfo {author} {\bibfnamefont {J.}~\bibnamefont
  {Steinhoff}},\ }\href {\doibase 10.1088/1475-7516/2014/12/003} {\bibfield
  {journal} {\bibinfo  {journal} {JCAP}\ }\textbf {\bibinfo {volume} {1412}},\
  \bibinfo {pages} {003} (\bibinfo {year} {2014})},\ \Eprint
  {http://arxiv.org/abs/1408.5762} {arXiv:1408.5762 [gr-qc]} \BibitemShut
  {NoStop}%
\bibitem [{\citenamefont {Levi}\ and\ \citenamefont
  {Steinhoff}(2016{\natexlab{b}})}]{Levi:2016ofk}%
  \BibitemOpen
  \bibfield  {author} {\bibinfo {author} {\bibfnamefont {M.}~\bibnamefont
  {Levi}}\ and\ \bibinfo {author} {\bibfnamefont {J.}~\bibnamefont
  {Steinhoff}},\ }\href@noop {} {\  (\bibinfo {year} {2016}{\natexlab{b}})},\
  \Eprint {http://arxiv.org/abs/1607.04252} {arXiv:1607.04252 [gr-qc]}
  \BibitemShut {NoStop}%
\bibitem [{\citenamefont {Zlochower}\ \emph {et~al.}(2005)\citenamefont
  {Zlochower}, \citenamefont {Baker}, \citenamefont {Campanelli},\ and\
  \citenamefont {Lousto}}]{Zlochower:2005bj}%
  \BibitemOpen
  \bibfield  {author} {\bibinfo {author} {\bibfnamefont {Y.}~\bibnamefont
  {Zlochower}}, \bibinfo {author} {\bibfnamefont {J.~G.}\ \bibnamefont
  {Baker}}, \bibinfo {author} {\bibfnamefont {M.}~\bibnamefont {Campanelli}}, \
  and\ \bibinfo {author} {\bibfnamefont {C.~O.}\ \bibnamefont {Lousto}},\
  }\href {\doibase 10.1103/PhysRevD.72.024021} {\bibfield  {journal} {\bibinfo
  {journal} {Phys. Rev.}\ }\textbf {\bibinfo {volume} {D72}},\ \bibinfo {pages}
  {024021} (\bibinfo {year} {2005})},\ \Eprint
  {http://arxiv.org/abs/gr-qc/0505055} {arXiv:gr-qc/0505055} \BibitemShut
  {NoStop}%
\bibitem [{\citenamefont {Marronetti}\ \emph {et~al.}(2008)\citenamefont
  {Marronetti}, \citenamefont {Tichy}, \citenamefont {Br{\"u}gmann},
  \citenamefont {Gonzalez},\ and\ \citenamefont
  {Sperhake}}]{Marronetti:2007wz}%
  \BibitemOpen
  \bibfield  {author} {\bibinfo {author} {\bibfnamefont {P.}~\bibnamefont
  {Marronetti}}, \bibinfo {author} {\bibfnamefont {W.}~\bibnamefont {Tichy}},
  \bibinfo {author} {\bibfnamefont {B.}~\bibnamefont {Br{\"u}gmann}}, \bibinfo
  {author} {\bibfnamefont {J.}~\bibnamefont {Gonzalez}}, \ and\ \bibinfo
  {author} {\bibfnamefont {U.}~\bibnamefont {Sperhake}},\ }\href {\doibase
  10.1103/PhysRevD.77.064010} {\bibfield  {journal} {\bibinfo  {journal} {Phys.
  Rev.}\ }\textbf {\bibinfo {volume} {D77}},\ \bibinfo {pages} {064010}
  (\bibinfo {year} {2008})},\ \Eprint {http://arxiv.org/abs/0709.2160}
  {arXiv:0709.2160 [gr-qc]} \BibitemShut {NoStop}%
\bibitem [{\citenamefont {Lousto}\ and\ \citenamefont
  {Zlochower}(2008)}]{Lousto:2007rj}%
  \BibitemOpen
  \bibfield  {author} {\bibinfo {author} {\bibfnamefont {C.~O.}\ \bibnamefont
  {Lousto}}\ and\ \bibinfo {author} {\bibfnamefont {Y.}~\bibnamefont
  {Zlochower}},\ }\href {\doibase 10.1103/PhysRevD.77.024034} {\bibfield
  {journal} {\bibinfo  {journal} {Phys. Rev.}\ }\textbf {\bibinfo {volume}
  {D77}},\ \bibinfo {pages} {024034} (\bibinfo {year} {2008})},\ \Eprint
  {http://arxiv.org/abs/0711.1165} {arXiv:0711.1165 [gr-qc]} \BibitemShut
  {NoStop}%
\bibitem [{\citenamefont {Zlochower}\ \emph {et~al.}(2012)\citenamefont
  {Zlochower}, \citenamefont {Ponce},\ and\ \citenamefont
  {Lousto}}]{Zlochower:2012fk}%
  \BibitemOpen
  \bibfield  {author} {\bibinfo {author} {\bibfnamefont {Y.}~\bibnamefont
  {Zlochower}}, \bibinfo {author} {\bibfnamefont {M.}~\bibnamefont {Ponce}}, \
  and\ \bibinfo {author} {\bibfnamefont {C.~O.}\ \bibnamefont {Lousto}},\
  }\href {\doibase 10.1103/PhysRevD.86.104056} {\bibfield  {journal} {\bibinfo
  {journal} {Phys. Rev.}\ }\textbf {\bibinfo {volume} {D86}},\ \bibinfo {pages}
  {104056} (\bibinfo {year} {2012})},\ \Eprint {http://arxiv.org/abs/1208.5494}
  {arXiv:1208.5494 [gr-qc]} \BibitemShut {NoStop}%
\bibitem [{Note1()}]{Note1}%
  \BibitemOpen
  \bibinfo {note} {Https://portal.xsede.org/sdsc-comet}\BibitemShut {NoStop}%
\bibitem [{\citenamefont {L{\"o}ffler}\ \emph {et~al.}(2012)\citenamefont
  {L{\"o}ffler}, \citenamefont {Faber}, \citenamefont {Bentivegna},
  \citenamefont {Bode}, \citenamefont {Diener}, \citenamefont {Haas},
  \citenamefont {Hinder}, \citenamefont {Mundim}, \citenamefont {Ott},
  \citenamefont {Schnetter}, \citenamefont {Allen}, \citenamefont
  {Campanelli},\ and\ \citenamefont {Laguna}}]{Loffler:2011ay}%
  \BibitemOpen
  \bibfield  {author} {\bibinfo {author} {\bibfnamefont {F.}~\bibnamefont
  {L{\"o}ffler}}, \bibinfo {author} {\bibfnamefont {J.}~\bibnamefont {Faber}},
  \bibinfo {author} {\bibfnamefont {E.}~\bibnamefont {Bentivegna}}, \bibinfo
  {author} {\bibfnamefont {T.}~\bibnamefont {Bode}}, \bibinfo {author}
  {\bibfnamefont {P.}~\bibnamefont {Diener}}, \bibinfo {author} {\bibfnamefont
  {R.}~\bibnamefont {Haas}}, \bibinfo {author} {\bibfnamefont {I.}~\bibnamefont
  {Hinder}}, \bibinfo {author} {\bibfnamefont {B.~C.}\ \bibnamefont {Mundim}},
  \bibinfo {author} {\bibfnamefont {C.~D.}\ \bibnamefont {Ott}}, \bibinfo
  {author} {\bibfnamefont {E.}~\bibnamefont {Schnetter}}, \bibinfo {author}
  {\bibfnamefont {G.}~\bibnamefont {Allen}}, \bibinfo {author} {\bibfnamefont
  {M.}~\bibnamefont {Campanelli}}, \ and\ \bibinfo {author} {\bibfnamefont
  {P.}~\bibnamefont {Laguna}},\ }\href@noop {} {\bibfield  {journal} {\bibinfo
  {journal} {Class. Quant. Grav.}\ }\textbf {\bibinfo {volume} {29}},\ \bibinfo
  {pages} {115001} (\bibinfo {year} {2012})},\ \Eprint
  {http://arxiv.org/abs/1111.3344} {arXiv:1111.3344 [gr-qc]} \BibitemShut
  {NoStop}%
\bibitem [{ein()}]{einsteintoolkit}%
  \BibitemOpen
  \href@noop {} {}\bibinfo {note} {Einstein Toolkit home page: {\tt
  http://einsteintoolkit.org}}\BibitemShut {NoStop}%
\bibitem [{cac()}]{cactus_web}%
  \BibitemOpen
  \href@noop {} {}\bibinfo {note} {Cactus Computational Toolkit home page: {\tt
  http://cactuscode.org}}\BibitemShut {NoStop}%
\bibitem [{\citenamefont {Schnetter}\ \emph {et~al.}(2004)\citenamefont
  {Schnetter}, \citenamefont {Hawley},\ and\ \citenamefont
  {Hawke}}]{Schnetter-etal-03b}%
  \BibitemOpen
  \bibfield  {author} {\bibinfo {author} {\bibfnamefont {E.}~\bibnamefont
  {Schnetter}}, \bibinfo {author} {\bibfnamefont {S.~H.}\ \bibnamefont
  {Hawley}}, \ and\ \bibinfo {author} {\bibfnamefont {I.}~\bibnamefont
  {Hawke}},\ }\href@noop {} {\bibfield  {journal} {\bibinfo  {journal} {Class.
  Quant. Grav.}\ }\textbf {\bibinfo {volume} {21}},\ \bibinfo {pages} {1465}
  (\bibinfo {year} {2004})},\ \Eprint {http://arxiv.org/abs/gr-qc/0310042}
  {gr-qc/0310042} \BibitemShut {NoStop}%
\bibitem [{\citenamefont {Thornburg}(2004)}]{Thornburg2003:AH-finding}%
  \BibitemOpen
  \bibfield  {author} {\bibinfo {author} {\bibfnamefont {J.}~\bibnamefont
  {Thornburg}},\ }\href {\doibase 10.1088/0264-9381/21/2/026} {\bibfield
  {journal} {\bibinfo  {journal} {Class. Quant. Grav.}\ }\textbf {\bibinfo
  {volume} {21}},\ \bibinfo {pages} {743} (\bibinfo {year} {2004})},\ \Eprint
  {http://arxiv.org/abs/gr-qc/0306056} {gr-qc/0306056} \BibitemShut {NoStop}%
\bibitem [{\citenamefont {Dreyer}\ \emph {et~al.}(2003)\citenamefont {Dreyer},
  \citenamefont {Krishnan}, \citenamefont {Shoemaker},\ and\ \citenamefont
  {Schnetter}}]{Dreyer02a}%
  \BibitemOpen
  \bibfield  {author} {\bibinfo {author} {\bibfnamefont {O.}~\bibnamefont
  {Dreyer}}, \bibinfo {author} {\bibfnamefont {B.}~\bibnamefont {Krishnan}},
  \bibinfo {author} {\bibfnamefont {D.}~\bibnamefont {Shoemaker}}, \ and\
  \bibinfo {author} {\bibfnamefont {E.}~\bibnamefont {Schnetter}},\ }\href@noop
  {} {\bibfield  {journal} {\bibinfo  {journal} {Phys. Rev.}\ }\textbf
  {\bibinfo {volume} {D67}},\ \bibinfo {pages} {024018} (\bibinfo {year}
  {2003})},\ \Eprint {http://arxiv.org/abs/gr-qc/0206008} {gr-qc/0206008}
  \BibitemShut {NoStop}%
\bibitem [{\citenamefont {Campanelli}\ \emph {et~al.}(2007)\citenamefont
  {Campanelli}, \citenamefont {Lousto}, \citenamefont {Zlochower},
  \citenamefont {Krishnan},\ and\ \citenamefont {Merritt}}]{Campanelli:2006fy}%
  \BibitemOpen
  \bibfield  {author} {\bibinfo {author} {\bibfnamefont {M.}~\bibnamefont
  {Campanelli}}, \bibinfo {author} {\bibfnamefont {C.~O.}\ \bibnamefont
  {Lousto}}, \bibinfo {author} {\bibfnamefont {Y.}~\bibnamefont {Zlochower}},
  \bibinfo {author} {\bibfnamefont {B.}~\bibnamefont {Krishnan}}, \ and\
  \bibinfo {author} {\bibfnamefont {D.}~\bibnamefont {Merritt}},\ }\href@noop
  {} {\bibfield  {journal} {\bibinfo  {journal} {Phys. Rev.}\ }\textbf
  {\bibinfo {volume} {D75}},\ \bibinfo {pages} {064030} (\bibinfo {year}
  {2007})},\ \Eprint {http://arxiv.org/abs/gr-qc/0612076} {gr-qc/0612076}
  \BibitemShut {NoStop}%
\bibitem [{\citenamefont {Campanelli}\ \emph
  {et~al.}(2006{\natexlab{b}})\citenamefont {Campanelli}, \citenamefont
  {Kelly},\ and\ \citenamefont {Lousto}}]{Campanelli:2005ia}%
  \BibitemOpen
  \bibfield  {author} {\bibinfo {author} {\bibfnamefont {M.}~\bibnamefont
  {Campanelli}}, \bibinfo {author} {\bibfnamefont {B.~J.}\ \bibnamefont
  {Kelly}}, \ and\ \bibinfo {author} {\bibfnamefont {C.~O.}\ \bibnamefont
  {Lousto}},\ }\href {\doibase 10.1103/PhysRevD.73.064005} {\bibfield
  {journal} {\bibinfo  {journal} {Phys. Rev.}\ }\textbf {\bibinfo {volume}
  {D73}},\ \bibinfo {pages} {064005} (\bibinfo {year} {2006}{\natexlab{b}})},\
  \Eprint {http://arxiv.org/abs/gr-qc/0510122} {arXiv:gr-qc/0510122}
  \BibitemShut {NoStop}%
\bibitem [{\citenamefont {Ansorg}\ \emph {et~al.}(2004)\citenamefont {Ansorg},
  \citenamefont {Br\"ugmann},\ and\ \citenamefont {Tichy}}]{Ansorg:2004ds}%
  \BibitemOpen
  \bibfield  {author} {\bibinfo {author} {\bibfnamefont {M.}~\bibnamefont
  {Ansorg}}, \bibinfo {author} {\bibfnamefont {B.}~\bibnamefont {Br\"ugmann}},
  \ and\ \bibinfo {author} {\bibfnamefont {W.}~\bibnamefont {Tichy}},\
  }\href@noop {} {\bibfield  {journal} {\bibinfo  {journal} {Phys. Rev.}\
  }\textbf {\bibinfo {volume} {D70}},\ \bibinfo {pages} {064011} (\bibinfo
  {year} {2004})},\ \Eprint {http://arxiv.org/abs/gr-qc/0404056}
  {gr-qc/0404056} \BibitemShut {NoStop}%
\bibitem [{\citenamefont {Bowen}\ and\ \citenamefont {York}(1980)}]{Bowen80}%
  \BibitemOpen
  \bibfield  {author} {\bibinfo {author} {\bibfnamefont {J.~M.}\ \bibnamefont
  {Bowen}}\ and\ \bibinfo {author} {\bibfnamefont {J.~W.}\ \bibnamefont {York},
  \bibfnamefont {Jr.}},\ }\href@noop {} {\bibfield  {journal} {\bibinfo
  {journal} {Phys. Rev.}\ }\textbf {\bibinfo {volume} {D21}},\ \bibinfo {pages}
  {2047} (\bibinfo {year} {1980})}\BibitemShut {NoStop}%
\bibitem [{\citenamefont {Cook}\ and\ \citenamefont
  {Pfeiffer}(2004)}]{Cook:2004kt}%
  \BibitemOpen
  \bibfield  {author} {\bibinfo {author} {\bibfnamefont {G.~B.}\ \bibnamefont
  {Cook}}\ and\ \bibinfo {author} {\bibfnamefont {H.~P.}\ \bibnamefont
  {Pfeiffer}},\ }\href@noop {} {\bibfield  {journal} {\bibinfo  {journal}
  {Phys. Rev. D}\ }\textbf {\bibinfo {volume} {70}},\ \bibinfo {pages} {104016}
  (\bibinfo {year} {2004})},\ \Eprint {http://arxiv.org/abs/gr-qc/0407078}
  {gr-qc/0407078} \BibitemShut {NoStop}%
\bibitem [{\citenamefont {Caudill}\ \emph {et~al.}(2006)\citenamefont
  {Caudill}, \citenamefont {Cook}, \citenamefont {Grigsby},\ and\ \citenamefont
  {Pfeiffer}}]{Caudill:2006hw}%
  \BibitemOpen
  \bibfield  {author} {\bibinfo {author} {\bibfnamefont {M.}~\bibnamefont
  {Caudill}}, \bibinfo {author} {\bibfnamefont {G.~B.}\ \bibnamefont {Cook}},
  \bibinfo {author} {\bibfnamefont {J.~D.}\ \bibnamefont {Grigsby}}, \ and\
  \bibinfo {author} {\bibfnamefont {H.~P.}\ \bibnamefont {Pfeiffer}},\
  }\href@noop {} {\bibfield  {journal} {\bibinfo  {journal} {Phys. Rev. D}\
  }\textbf {\bibinfo {volume} {74}},\ \bibinfo {pages} {064011} (\bibinfo
  {year} {2006})},\ \Eprint {http://arxiv.org/abs/gr-qc/0605053}
  {gr-qc/0605053} \BibitemShut {NoStop}%
\bibitem [{\citenamefont {Lovelace}\ \emph {et~al.}(2008)\citenamefont
  {Lovelace}, \citenamefont {Owen}, \citenamefont {Pfeiffer},\ and\
  \citenamefont {Chu}}]{Lovelace:2008tw}%
  \BibitemOpen
  \bibfield  {author} {\bibinfo {author} {\bibfnamefont {G.}~\bibnamefont
  {Lovelace}}, \bibinfo {author} {\bibfnamefont {R.}~\bibnamefont {Owen}},
  \bibinfo {author} {\bibfnamefont {H.~P.}\ \bibnamefont {Pfeiffer}}, \ and\
  \bibinfo {author} {\bibfnamefont {T.}~\bibnamefont {Chu}},\ }\href {\doibase
  10.1103/PhysRevD.78.084017} {\bibfield  {journal} {\bibinfo  {journal} {Phys.
  Rev.}\ }\textbf {\bibinfo {volume} {D78}},\ \bibinfo {pages} {084017}
  (\bibinfo {year} {2008})},\ \Eprint {http://arxiv.org/abs/0805.4192}
  {arXiv:0805.4192 [gr-qc]} \BibitemShut {NoStop}%
\bibitem [{\citenamefont {Ruchlin}\ \emph {et~al.}(2016)\citenamefont
  {Ruchlin}, \citenamefont {Healy}, \citenamefont {Lousto},\ and\ \citenamefont
  {Zlochower}}]{Ruchlin:2014zva}%
  \BibitemOpen
  \bibfield  {author} {\bibinfo {author} {\bibfnamefont {I.}~\bibnamefont
  {Ruchlin}}, \bibinfo {author} {\bibfnamefont {J.}~\bibnamefont {Healy}},
  \bibinfo {author} {\bibfnamefont {C.~O.}\ \bibnamefont {Lousto}}, \ and\
  \bibinfo {author} {\bibfnamefont {Y.}~\bibnamefont {Zlochower}},\ }\href@noop
  {} {\  (\bibinfo {year} {2016})},\ \bibinfo {note} {accepted to Phys. Rev. D
  (Dec. 2016)},\ \Eprint {http://arxiv.org/abs/1410.8607} {arXiv:1410.8607
  [gr-qc]} \BibitemShut {NoStop}%
\bibitem [{\citenamefont {Carroll}(2004)}]{Carroll:2004st}%
  \BibitemOpen
  \bibfield  {author} {\bibinfo {author} {\bibfnamefont {S.~M.}\ \bibnamefont
  {Carroll}},\ }\href
  {http://www.slac.stanford.edu/spires/find/books/www?cl=QC6:C37:2004} {\emph
  {\bibinfo {title} {{Spacetime and geometry: An introduction to general
  relativity}}}}\ (\bibinfo  {publisher} {Addison-Wesley},\ \bibinfo {year}
  {2004})\BibitemShut {NoStop}%
\bibitem [{\citenamefont {Yunes}\ and\ \citenamefont
  {Berti}(2008)}]{Yunes:2008tw}%
  \BibitemOpen
  \bibfield  {author} {\bibinfo {author} {\bibfnamefont {N.}~\bibnamefont
  {Yunes}}\ and\ \bibinfo {author} {\bibfnamefont {E.}~\bibnamefont {Berti}},\
  }\href {\doibase 10.1103/PhysRevD.77.124006, 10.1103/PhysRevD.83.109901}
  {\bibfield  {journal} {\bibinfo  {journal} {Phys. Rev.}\ }\textbf {\bibinfo
  {volume} {D77}},\ \bibinfo {pages} {124006} (\bibinfo {year} {2008})},\
  \Eprint {http://arxiv.org/abs/0803.1853} {arXiv:0803.1853 [gr-qc]}
  \BibitemShut {NoStop}%
\bibitem [{\citenamefont {Zhang}\ \emph {et~al.}(2011)\citenamefont {Zhang},
  \citenamefont {Yunes},\ and\ \citenamefont {Berti}}]{Zhang:2011vha}%
  \BibitemOpen
  \bibfield  {author} {\bibinfo {author} {\bibfnamefont {Z.}~\bibnamefont
  {Zhang}}, \bibinfo {author} {\bibfnamefont {N.}~\bibnamefont {Yunes}}, \ and\
  \bibinfo {author} {\bibfnamefont {E.}~\bibnamefont {Berti}},\ }\href
  {\doibase 10.1103/PhysRevD.84.024029} {\bibfield  {journal} {\bibinfo
  {journal} {Phys. Rev.}\ }\textbf {\bibinfo {volume} {D84}},\ \bibinfo {pages}
  {024029} (\bibinfo {year} {2011})},\ \Eprint {http://arxiv.org/abs/1103.6041}
  {arXiv:1103.6041 [gr-qc]} \BibitemShut {NoStop}%
\bibitem [{\citenamefont {Sago}\ \emph {et~al.}(2016)\citenamefont {Sago},
  \citenamefont {Fujita},\ and\ \citenamefont {Nakano}}]{Sago:2016xsp}%
  \BibitemOpen
  \bibfield  {author} {\bibinfo {author} {\bibfnamefont {N.}~\bibnamefont
  {Sago}}, \bibinfo {author} {\bibfnamefont {R.}~\bibnamefont {Fujita}}, \ and\
  \bibinfo {author} {\bibfnamefont {H.}~\bibnamefont {Nakano}},\ }\href
  {\doibase 10.1103/PhysRevD.93.104023} {\bibfield  {journal} {\bibinfo
  {journal} {Phys. Rev.}\ }\textbf {\bibinfo {volume} {D93}},\ \bibinfo {pages}
  {104023} (\bibinfo {year} {2016})},\ \Eprint
  {http://arxiv.org/abs/1601.02174} {arXiv:1601.02174 [gr-qc]} \BibitemShut
  {NoStop}%
\end{thebibliography}%

\end{document}